\documentclass{aastex63}

\usepackage{graphicx}
\usepackage{float}
\usepackage{subfigure}
\shortauthors{Zhuotao Li et al.}

\begin{document}
	\title{Reconstruction of Supernova Gravitational Waves Waveforms: Comparing Three Time-frequency Transform Methods}
		\correspondingauthor{Xilong Fan}
	\email{xilong.fan@whu.edu.cn}
	
	\author{Zhuotao Li}
	\affiliation{School of Physics Science And Technology, Wuhan University,
		No.299 Bayi Road, Wuhan, Hubei, China}
	
	\author{Xilong Fan}
	\affiliation{School of Physics Science And Technology, Wuhan University,
		No.299 Bayi Road, Wuhan, Hubei, China}
	
	\author{Gang Yu}
\affiliation{School of Electrical Engineering, University of Jinan, Jinan 250022, China}
	
	\begin{abstract}
		For supernovae gravitational wave signal analysis which intend to reconstruct   supernova gravitational waves waveforms, we compare the performance of short-time Fourier transform (STFT), the synchroextracting transform (SET) and multisynchrosqueezing transform (MSST)  by a self-consistent  time-frequency analysis based  pipeline.  The simulated supernovae waveforms injected into white noise are identified by a  hierarchical clustering method in the time-frequency map and then reconstructed by the inverse time-frequency transforms. We find that in terms of signal reconstruction, the SET method performed the best, especially much better than traditional STFT method in reconstructing signals from data with white noise when valued the signal-to-noise ratio.  While concerning the quality of time-frequency figures, the MSST method and SET method have less energy dispersion and were both better than STFT method. The higher  energy dispersion in  time-frequency figure of STFT is  time consuming in the  clustering  process and reduce the accuracy of   signal identification. Our preliminary conclusion is that the SET method is the  suitable method for the supernovae gravitational wave signal analysis pipeline though more tests are stilled needed.
\end{abstract}
	\keywords{gravitational wave, supernovae, time-frequency analysis}
	
	\section{Introduction}
	In 2015, LIGO observed a gravitational wave signal for the first time \citep{PhysRevLett.116.061102}. Since then gravitational wave observation has been applied to a number of researches on high-energy  celestial bodies and events including binary black hole mergers, binary neutron star mergers, and core-collapse supernovae. For high-energy celestial body mergers, the physical process is rather clear and the waveform template is calculable \citep{PhysRevLett.116.061102}. However, the physics of core-collapse supernovae is still not clear and has become an important topic.
	
	The mechanism of core-collapse supernovae explosion has long been discussed. When a massive star (10 - 100 $M_\odot$) reaches the final stage of its life, core-collapse will happen until its density is comparable to that of nuclear matter \citep{Ott_2009}. The inner core will then bounce back and form a shock wave, which will soon stall due to dissociation of iron nuclei and neutrino cooling. What mechanism stores the energy of the first shock wave and revives it is the key to the mechanism of supernovae \citep{KOTAKE2013318}. The two most prevailing theories are neutrino heating mechanism \citep{1966ApJ...143..626C} and magnetorotational mechanism \citep{Woosley_2006}. The former of which relies on neutrino heating process to store the energy in neutrinos to revive the shock wave, suited for non-rotating or slowly rotating massive stars. At the same time, the later of which transforms the rotational kinetic energy to the energy of the shock wave, suited for rapidly rotating massive stars  \citep{Ott_2009}.
	
	However, the verification of the two theories is not a simple task since these dynamics take place inside the stellar core, shrouded from our view. Besides, the observation for supernovae neutrinos will also face the problem that the neutrinos will be affected by the Mikheyev–Smirnov–Wolfenstein effect in propagating the stellar envelope. In this case, the most practical solution is to observe supernovae gravitational wave (SNGW), which can carry information directly to us without being affected \citep{KOTAKE2013318}. With the help of current ground-based gravitational wave detectors, it is now possible to detect CCSN sources within the distance of a few kpcs. The question is how to distinguish an SNGW signal from noise. The most prevalent way is to apply time-frequency analysis which is also used in the discovery of gravitational waves to find the 'chirp' signal \citep{PhysRevLett.116.061102}. However, the resolution of the traditional time-frequency transform, short-time Fourier transform (STFT), is limited, and its dispersion increases the computation greatly, which may become an obstacle for the clustering algorithm.
	
	In recent years, with the develop of time-frequency analysis, some new techniques with better resolution such as synchroextracting transform (SET) \citep{7906573} and multisynchrosqueezing transform (MSST) \citep{8458385} have appeared. Time-frequency analysis (TFA) techniques characterize the time-varying structures of the signals by calculating the inner product between them and a series of time-frequency (TF) atoms. However, the TF results created by traditional TFA techniques suffer from energy dispersion problem due to the limitation of the Heisenberg uncertainty principle. This also leads to the poor TF resolution in the analysis of multi-component signals. Various TFA techniques are recently proposed for solving such a problem, such as SET and MSST. SET employs a Dirac delta operator to remove the redundant TF information. MSST is proposed to solve the energy dispersion problem using the squeezing strategy. Both two techniques have the capacity to improve the TF resolution in the analysis of multi-component signals. However, it is an open issue that which one is more suitable for addressing SNGW signals.
	
	Another problem is that how to distinguish signals once we have the time-frequency figure. The  traditional simple clustering method classifies neighbouring pixels into one cluster, meanwhile the dispersion of the STFT method turns the signals and noise into a 2D region in the figure. Due to this reason, signals can be corrupted by nosie during the clustering step. Thus, we applied hierarchical algorithm which computes the SNR of clusters and will be introduced later.
	
	In this paper, we constructed a time-frequency analysis pipeline and tested the performances of these STFT, SET and MSST with simulated SNGW data. In addition, we also applied hierarchical algorithm \citep{Heng_2004} in clustering instead of the traditional simple clustering method. We mainly focus on the quality of the time-frequency figure and reconstruction of signals.
	
	Section II will give a brief introduction to the simulated data we used. In Section III, we will introduce the three time-frequency transform methods, respectively. A general pipeline will be described in Section IV. In Section V, we will show the results of the tests and compare the advantages and disadvantages of the three methods. A discussion of reconstruction schemes will also be given. In Section VI, we summarize our work.
	
	\section{Simulated Data}
	The simulated data we used in this paper consists of two parts, simulated signals and simulated white noise into which we inject the signals. Each set of white noise is generated randomly and is of the same average amplitude as that of the injected signal. As for the main features of simulated SNGW signals as well as signals used in this paper, they will be introduced in the following.
	
	For neutrino heating mechanism which is suited for slowly rotating massive stars, SASI and convection usually lead to stochastic, broad-band signals. For massive stars that are rapidly rotating, the magnetorotational mechanism often leads to a bounce and a tail in broad-band signals in the 2D simulation due to the core bounce and magnetohydrodynamic outflows, and sometimes long-lasting narrow-band signals in the 3D simulation due to non-axisymmetric rotational instabilities \citep{KOTAKE2013318}.	
	\begin{figure}[htbp]
	\centering
	\subfigure[Abd-2014-A1O03.5 \cite{Abdikamalov_2014}, 2D magnetorotational]{
		\includegraphics[width=0.45\textwidth]{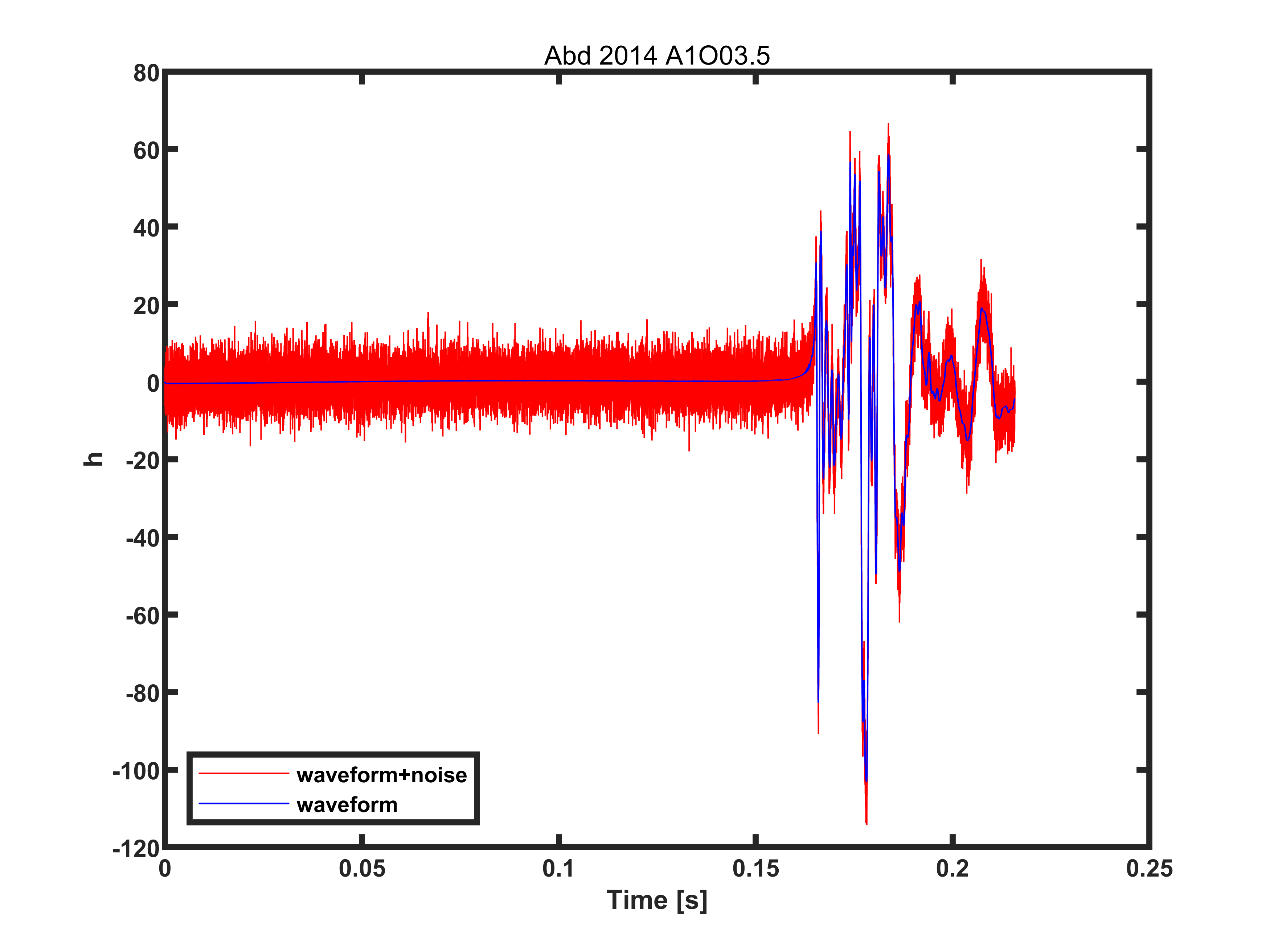}}
	\subfigure[R1E1CA-L \cite{Scheidegger_2010}, 3D magnetorotational]{
		\includegraphics[width=0.45\textwidth]{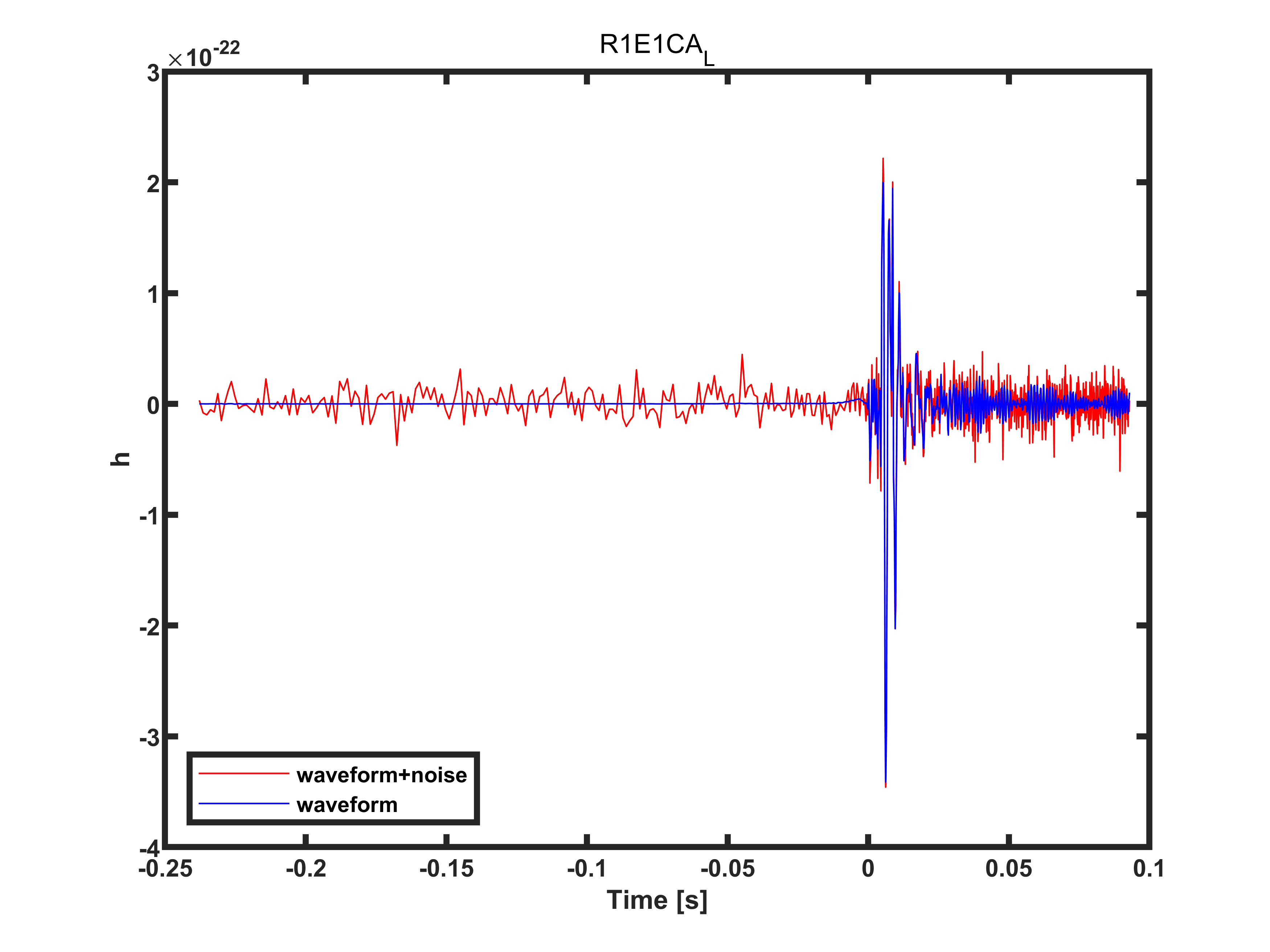}}
	\subfigure[C15-3D GWsignal 0-438ms \cite{yakunin2017gravitational}, 3D neutrino-driven]{
		\includegraphics[width=0.45\textwidth]{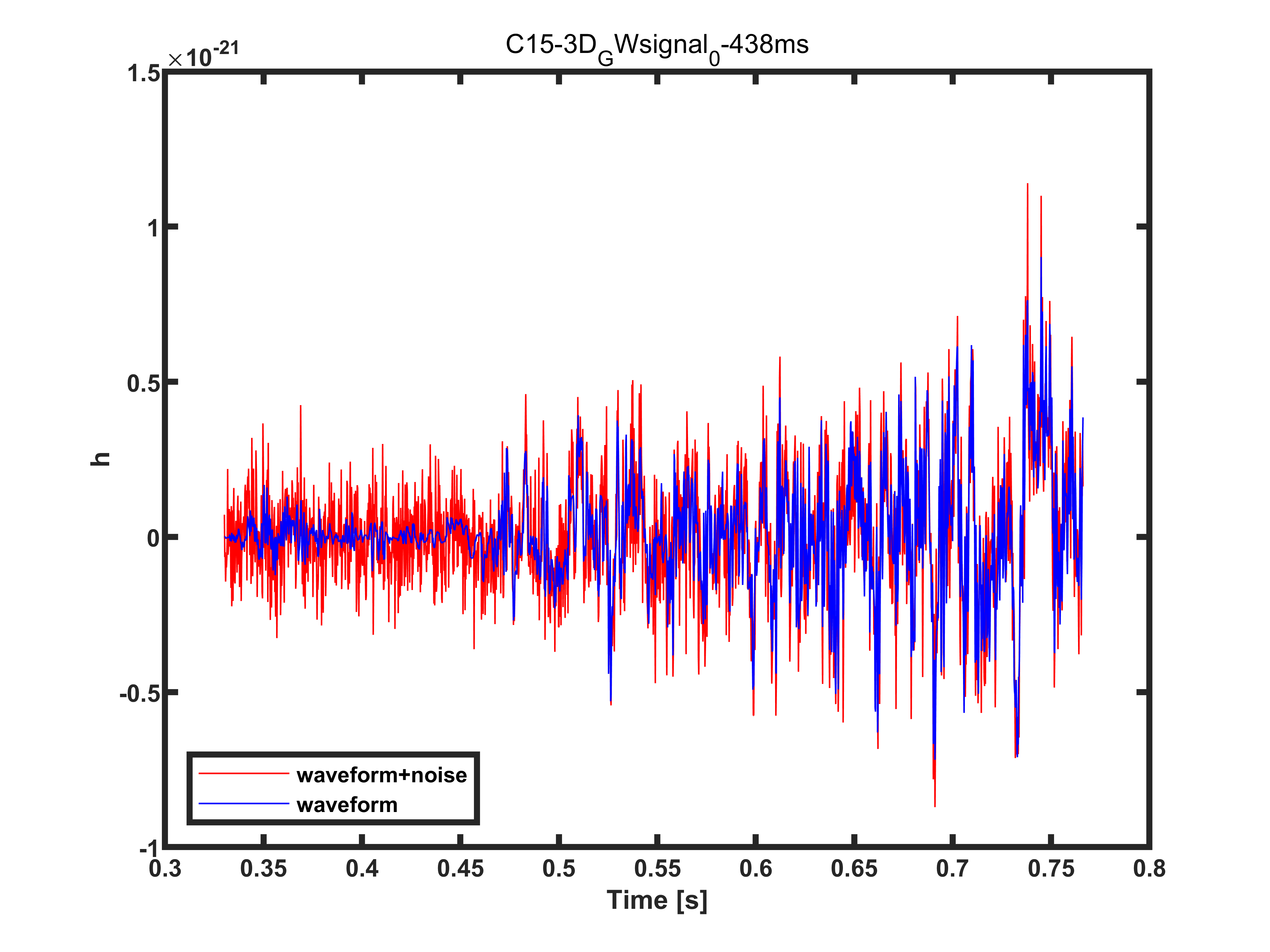}}
	\caption{These are three examples of the signals we used, the corresponding mechanism and dimension are noted. The x-axis represents time while y-axis represents the waveform. Red curves represents data of injected signals and white noise which is of the same average amplitude as the signal while blue curves represents pure signals.}
    \end{figure}
	Based on the features of SNGW signals, the time-frequency transform methods we need should have the ability to find the low-frequency mode and check if there is a bounce or tail. It should also illustrate the information of the high-frequency band. However, it is worth mentioning that distinguishing high-frequency stochastic signals from noise is not simple; this further adds to the difficulty of SNGW signal processing.

After we have the results of the signal processing, our criteria are as follows. A low-frequency mode without a bounce plus relatively small high-frequency stochastic signals should be generated from the neutrino mechanism. If a signal has a low-frequency mode with a bounce or only have high-frequency components with no low-frequency mode, we can almost be certain that it is from the magnetorotational mechanism.  In our research, the data we use is generated by whited noise. Table \ref{snsignals} shows the references of the simulated GW waveforms.
	\begin{table}[h]
		\caption{Simulated Signals}\label{snsignals}
		\begin{tabular}{|l|l|l|l|l|}
			\hline
			Author & Year & Dimension & Mass & Reference
			\\
			\hline
			Abdikamalov & 2014 & 2D & 12$M_\odot$ & \cite{Abdikamalov_2014}
			\\
			\hline
			Andresen & 2016 & 3D & 11.2, 20 and 27$M_\odot$ & \cite{Andresen_2017}
			\\
			\hline
			Andresen & 2019 & 3D & 15$M_\odot$ & \cite{Andresen_2019}
			\\
			\hline
			Cerda-Duran & 2013 & 2D & 35$M_\odot$ & \cite{Cerd_Dur_n_2013}
			\\
			\hline
			Dimmelmeier & 2008 & 2D & 10-100$M_\odot$ & \cite{Dimmelmeier_2008}
			\\
			\hline
			Kuroda2016 & 2016 & 3D & 15$M_\odot$ & \cite{Kuroda_2016}
			\\
			\hline
			Kuroda & 2017 & 3D & 11, 15$M_\odot$ & \cite{Kuroda_2017}
			\\
			\hline
			Morozova & 2018 & 3D & 10, 13 and 19$M_\odot$ & \cite{Morozova_2018}
			\\
			\hline
			Mueller & 2012 & 3D & 15, 20$M_\odot$ & \cite{M_ller_2012}
			\\
			\hline
			Oconnor Couch & 2018 & 2D,3D & 20$M_\odot$ & \cite{2018ApJ...865...81O}
			\\
			\hline
			Ott & 2013 & 3D & 27$M_\odot$ & \cite{Ott_2013}
			\\
			\hline
			Powell & 2018 & 3D & 3.5, 18$M_\odot$ & \cite{Powell_2019}
			\\
			\hline
			Radice & 2019 & 3D & 9, 10, 11, 12, 13, 19, 25 and 60$M_\odot$ & \cite{Radice_2019}
			\\
			\hline
			Scheidegger & 2010 & 3D & 15$M_\odot$ & \cite{Scheidegger_2010}
			\\
			\hline
			Yakunin & 2015 & 2D & 12, 15, 20 and 25$M_\odot$ & \cite{yakunin2015gravitational}
			\\
			\hline
			Yakunin & 2017 & 3D & 15$M_\odot$ & \cite{yakunin2017gravitational} 
			\\
			\hline
		\end{tabular}
	\end{table}

	\section{Time-frequency transform and corresponding inverse transform}
	In this paper, three time-frequency transform methods are used, which are STFT (short-time Fourier transform), SET (synchroextracting transform) \citep{7906573}, MSST (multisynchrosqueezing transform) \citep{8458385} The advantages and disadvantages are compared. For the three time-frequency transform, STFT, SET, MSST, their inverse transforms will also be introduced.
	\subsection{STFT}
	Short-time Fourier transform is the most classical time-frequency transform, which was proposed by Dennis Gabor in 1946. The basic idea is to use the window function to intercept the signal and assume that it is stationary in the window, and then do Fourier transform to the signal in the window. Next, the window function is translated along the time direction, and the Fourier transform results at different time positions are put together to obtain the frequency change relationship with time. The STFT formula of signal $x(u)$ is as follows:
	\begin{equation}
	G(t, \omega)=\int_{-\infty}^{+\infty} g(u-t) s(u) e^{-i \omega(u-t)} d u,
	\end{equation}		
where $x(T)$ is the signal to be analyzed, $g(T)$ is the window function. Its inverse transformation is
	\begin{equation}
	s(t)=(2 \pi g(0))^{-1} \int_{-\infty}^{+\infty} G(t, \omega) d \omega
	\end{equation}		
	\subsection{SET}
	In theory, STFT calculates the Fourier transform of the signal in a short time truncated by the window. Although the STFT can characterize the time-varying features of a non-stationary signal, the TF representation (TFR) smears in a large region since the window cannot compactly support in both time and frequency domain. This must lead to the energy dispersion problem. To concentrate the energy, SET is proposed using a Dirac delta function as described in (3).
	\begin{equation}
	T e(t, \omega)=G(t, \omega) \delta(\omega-\hat{\omega}(t, \omega)),
	\end{equation}
	where $\hat{\omega}$ is termed the two-dimensional instantaneous frequency of the STFT as calculated in (4).
	\begin{equation}
	\hat{\omega}(t, \omega)=\frac{\partial_{t} G(t, \omega)}{i G(t, \omega)}.
	\end{equation}
	The Dirac delta function employed in the SET algorithm can remove most of the redundant TF coefficients caused by the energy dispersion problem. Thus, SET has the capacity to provide an energy-concentrated TFR for describing the time-varying features of the non-stationary signals. Moreover, the SET allows for an approximated reconstruction for the signal, as expressed in (5).
    $$
    \left.s(t) \approx T e(t, \omega)\right|_{\omega=\varphi^{\prime}(t)},
    $$
where $\varphi^{\prime}(t)$ is the TF ridge of the signal in the TFR of the SET. For more details please check \cite{7906573}.
	
	\subsection{MSST}
	Although the SET has better resolution than the STFT, it cannot provide a perfect reconstruction for the signal. This may miss some information in the analysis of the signal affected by heavy noise. To solve this problem, MSST is proposed by employing an iterative reassignment operator for concentrating the TF energy of the STFT as written in (5).
	\begin{equation}
	\begin{array}{l}
	T s^{[1]}(t, \eta)=\int_{-\infty}^{+\infty} G(t, \omega) \delta(\eta-\hat{\omega}(t, \omega)) d \omega \\
	T s^{[2]}(t, \eta)=\int_{-\infty}^{+\infty} T s^{[1]}(t, \omega) \delta(\eta-\hat{\omega}(t, \omega)) d \omega \\
	T s^{[3]}(t, \eta)=\int_{-\infty}^{+\infty} T s^{[2]}(t, \omega) \delta(\eta-\hat{\omega}(t, \omega)) d \omega \\
	\vdots \\
	T s^{[N]}(t, \eta)=\int_{-\infty}^{+\infty} T s^{[N-1]}(t, \omega) \delta(\eta-\hat{\omega}(t, \omega)) d \omega
	\end{array}
	\end{equation}
	
	Because MSST only considers the TF reassignment in the frequency direction, the signal can be perfectly reconstructed by (6).
	\begin{equation}
	s(t)=(2 \pi g(0))^{-1} \int_{-\infty}^{+\infty} T s^{[N]}(t, \omega) d \omega
	\end{equation}
	For more details on MSST please check \cite{8458385}.
	
	\section{PIPELINE}
	In this section, we will construct the basic pipeline of our tests though some of the steps will be skipped, which will be discussed later. Our main method is to use time-frequency transform and the hierarchical clustering method to distinguish and reconstruct the signal.
	
	
	The final goal of the time-frequency analysis of the core-collapse supernovae GW signals is to reconstruct and analyze the original signal to determine the explosion mechanism. To achieve this goal, the pipeline is divided into five steps:
	
	1) Whiten the noise to convert the characteristic noise in the data into white noise, which is convenient for subsequent processing.
	
	2) Suppress white noise.
	
	
	3) Time-frequency transform of the signal.
	
	4) Use the hierarchical clustering method to identify the position of the signal, then analyze the time-frequency diagram of the signal.
	
	5) Reconstruct the signal using the inverse time-frequency transform of the identified signal.
	
	After this process, we are able to analyze the reconstructed signal, find its characteristics, and then determine the mechanism of the supernova explosion.
	
	These steps will be introduced in the rest of this part except the time-frequency transform as well as the inverse transform which are introduced in part III. The core of this process is the time-frequency transform method. We tested three methods, namely SET, MSST, and the traditional STFT, with simulated signals.

	\subsection{Whitening}
	The GW signals of supernovae generally contain noises with various sources which can be divided into the white noise and the characteristic noise. To eliminate the characteristic noise in the frequency band and transform it into white noise, we use the PSD obtained in the previous step and further extract ASD (amplitude spectral density) to whiten the original data in the time domain and make the noise covariance matrix to be 1. Let X be a random signal vector with zero mean value, and its autocorrelation matrix is
	$$R_x=E[xx^T],$$
where $R_x$ is a symmetric matrix and nonnegative definite. Now find a linear transformation $B$ to transform $x$, that is, $y = BX$, so that
	$$R_y=BE[xx^T]B^T ,$$
where $B$ is the whitening matrix, the meaning of the above formula is: the components of $Y$ are irrelevant. This process is often referred to as "spatial decorrelation" or "spatial whitening".
	The formula for using ASD to whiten the noise is as follows:
	\begin{equation}
	d f=\frac{2 H(f) d t}{A(f)}
	\end{equation}
	
	Among them, $df$ is the frequency domain data after noise whitening, $H(f)$ is the frequency domain data of original data after Fourier transform, $dt$ is the time interval between two adjacent points, and $A(f)$ is ASD, i.e. PSD square root. However, because we only tested simulated SNe GW signals with no characteristic noise since SNe GW signals have not been detected yet, we skipped this step in our test. But it is surely useful when dealing with a real SNe GW signal.
	
	\subsection{Noise reduction}
	There are two main problems of noise reduction when dealing with SNe GW signals. First, there is no common template of the signal \citep{KOTAKE2013318}; people merely know the characteristics of different explosion mechanisms. Thus the difficulty of detecting the signal from white noise will be greater than other sources like neutron star mergers. Currently, there is no analytic solution of SNe GW signals, and the numerical solutions differed from each other depending on the model even if the mechanism is the same. Secondly, the SNe GW signals are assumed to be stochastic, especially in the neutrino-driven mechanism, which makes it difficult to distinguish the signal from the noise if the average amplitude of the noise is much higher than that of the signal. The process of smoothing the signal can also let the signal lose some of its high-frequency band information, which can be important in determining the explosion mechanism. This further adds the difficulty for noise reduction.
	
	For the two reasons above, we only applied a simple method with light effect on the data to mitigate the effect on the signal. The method is taking the average value in a bin as the value at the middle point to reduce the white noise since the average value of the white noise should approach 0 in a long time.
	\begin{equation}
	X_{\text{denoised}}(t)=\frac{\int_{t-\tau / 2}^{t+\tau / 2} X_{\text{original}}(a) d a}{\tau},
	\end{equation}
where $X_{denoised}(t)$ is the result, $X_{original}(a)$ is the original signal, and $\tau$ is the length of a bin.

%
%
%
%
%

	\subsection{Hierarchical clustering}
	Through time-frequency transform an $N*T$ matrix is obtained. If it is put on a two-dimensional plane, the X-axis will represent time. There are a total of $T$ columns, representing t from the beginning of time $T_1$ to $T_ {end}=T_ 1 + (T-1) * dT$; the y-axis represents the frequency, with $N$ columns in total, representing the frequency from $df$ to f from the beginning $f_end=df*(N-1)$. The relative magnitude of the modulus length of the complex number at each position represents the relative power of the frequency at that time.
	
	To recognize the signal from noise, we adopted the method of hierarchical clustering scheme instead of the simple clustering method which classifies neighbouring pixels into one cluster, for more details please check \cite{Heng_2004}. The main idea is to calculate the SNR (signal-to-noise ratio) of every TF pixel, set an SNR threshold, find all pixels above this threshold, identify clusters of TF pixels, then find the clusters above the second SNR threshold. These clusters that survived the second threshold are considered as signals.

	\subsubsection{Compute SNR}
	First, we calculate the mean value and RMS of the all pixels except those whose value is 10\% extreme to avoid the extremes of the distribution, which could contain high SNR events, from biasing the estimate of the mean and RMS, in every frequency bin. Then all pixels in this frequency bin are divided by the corresponding RMS value; the results are SNR values of every pixel. A threshold will then be applied to all the pixels; the value of the threshold is decided by experience; in this paper, we use 10\% of the highest SNR value. Pixels with SNR lower than the threshold will be removed. Next, the remaining pixels will be clustered.
	
	
	\subsubsection{Clustering}
	The mathematical principle of clustering is that each sample object, in this paper, the matrix element of $N*T$ matrix, is regarded as a point in the high-dimensional space, and then through a certain way to define and calculate the distance between sample points in high-dimensional space, and classify the closer points into a class, and calculate the distance between classes.
	In this paper, in a simple case, the matrix elements of the $N*T$ matrix filtered by a threshold can be directly placed in the time-frequency two-dimensional space, ignoring the difference in the size of its data modulus; or it can be processed in three-dimensional space, each dimension represents frequency, time, and data modulus respectively.
	
	Before calculating the distance in a high-dimensional space, we first need to centralize and standardize the attribute indexes of each dimension to eliminate errors caused by different attribute index scales. The standardized method is
	\begin{equation}
	X_{i}^{\prime}=\frac{X_{i}-\bar{X}}{\sigma},
	\end{equation}
	where $\bar{X}$ is the mean value, and $\sigma$ is the standard deviation. Then the distance in a high-dimensional space can be calculated, and commonly used methods for calculating the distance between sample points are as follows.
	
    \begin{table}[htbp]
    	\caption{Methods of Calculating Distances}
    	\begin{tabular}{|l|l|}
    		\hline
    		Absolute Distance & $d_{rs}=\sum_{j=1}^M{\vert z_{rj}-z_{sj}\vert}$ \\
    		\hline
    		Euclidean Distance & $d_{rs}(2)=[\sum_{j=1}^M{( z_{rj}-z_{sj})^2}]^\frac{1}{2}$
    		\\
    		\hline
    		Standard Euclidean Distance & Euclidean distance with reciprocal of sample variance as weight
    		\\
    		\hline
    		Minkowski Distance & $d_{rs} (p)=[\sum_{j=1}^M (z_{rj}-z_{sj} )^p]^\frac{1}{p}$
    		\\
    		\hline
    		Chebyshev Distance & $d_{rs}({dim})={max}_{1\le j\le {dim}} \vert z_{rj}-z_{sj}\vert $
    		\\
    		\hline
    		Mahalanobis Distance & $d_{rs}^2=(z_r-z_s ) V^{-1} (z_r-z_s )^T $ V is the covariance matrix
    		\\
    		\hline
    	\end{tabular}
    \end{table}
	
	After the above methods obtain the distances of the sample points, they can be classified, and the relatively close points are classified as a category. However, for the SET method and MSST method, the time-frequency figure is composed of independent and consecutive curves. For these two methods, we recognize every curve as a cluster and take relatively long curves as signals. As for the STFT method, the computation of distances among the points is usually too great to carry out because the dispersion increased the number of sample points greatly.
	
	After clustering,  the highest pixel SNR in every cluster will be considered as the SNR of the cluster. Then a second threshold will be applied to all the clusters. The value of the threshold is also by experience, and in this paper, we use 10\% of the highest SNR value. Those clusters which survived the second threshold will be regarded as signals \citep{Heng_2004}. It is also worth mentioning that, when the noise amplitude is much smaller than the average amplitude of the signal, filtering and clustering have little effect on or even worsen the results, because they can leak the energy of the signal.

	\section{Results}
	We have tested simulated data which is simulated signals injected into whited noise which is of the same average amplitude of that signal from part II with the pipeline and evaluated the results with the signal-to-noise ratio between the simulated data and the reconstructed signal. Further more, in the next step of our research, we are planing to inject the signals into real LIGO noise. The results differed depending on the time-frequency transform methods.
However, due to limited computational power and RAM, we had to take three measures to decrease computation. First, we divided the data into many bins to treat each bin separately. However, it would damage the quality of reconstruction signals of the MSST method, as we will mention later.Second, since simulated signals had no characteristic noise, we skipped whitening. Third, for the STFT method, the computation of distances was too great and thus We had to skip clustering due to the limitations of computing power and time for it.
	
	The results will be valued from two aspects, the quality of f-t figures and the reconstruction of signals, whose quality would be valued by the signal-to-noise ratio between the template and the reconstructed signals. The steps of computing SNR between data $s(n)$ and template $h(n)$ of length N are: first, compute the single-sided FFT (Fast-Fourier Transform) of $s(n)$ and $h(n)$, the results are $s(f_k)$ and $h(f_k)$; second, estimate noise power density spectrum (PSD) $P_{ww}(f_k)$; third, compute the matched filter output $z(n)$ and normalization factor $\rho_h$:
	\begin{equation}
	z(n)=4 \sum_{k=1}^{\frac{N-1}{2}} \frac{s\left(f_{k}\right) h^{*}\left(f_{k}\right)}{P_{w w}\left(f_{k}\right)} e^{i 2 \pi n k / N}
	\end{equation}
	\begin{equation}
	\rho_{h}^{2}=4 \Delta f \sum_{k=1}^{\frac{N-1}{2}} \frac{\left|h\left(f_{k}\right)\right|^{2}}{P_{w w}\left(f_{k}\right)}
	\end{equation}
Last, compute the signal-to-noise ratio:
	\begin{equation}
	\rho(n)=\frac{|z(n)|}{\rho_{h}}
	\end{equation}
	The greatest value in the array $\rho(n)$, $\max{\rho(n)}$ is the predicted SNR. A more detailed description is in \cite{AntelisMoreno-41}.

	\subsection{Time-frequency figure}
	The time-frequency figure of data consisting signal "C15-3D GWsignal 0-438ms" and white noise of the same average amplitude by three time-frequency transform methods is:
	\begin{figure}[h]
		\centering
		\includegraphics[width=0.9\textwidth]{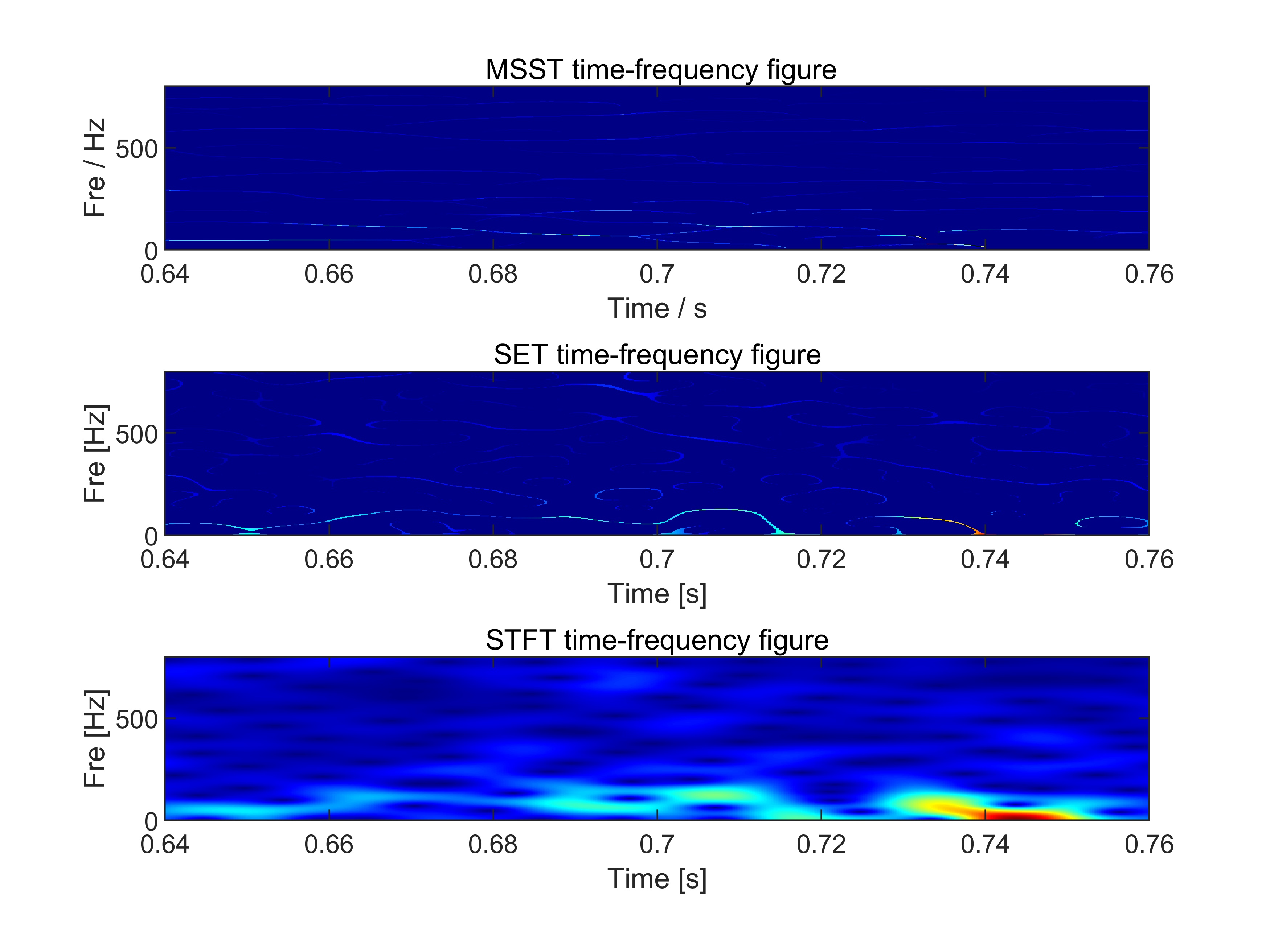}
		\caption{These are the results of the three methods for the signal C15-3D GWsignal 0-438ms. The x-axis represents time while y-axis represents frequency. The color at the point (T,F) represents the power of the signal at time T and frequency F. The brighter the color is, the greater the power is.}
	\end{figure}
	The signal and noise in the figure of the STFT method are represented by imprecise bright areas while by precise bright curves in the figure of the SET method and the MSST method. . This shows the defect of the traditional STFT method, great dispersion, leading to unsatisfactory f-t figure results. However, the performances of MSST and SET are much better, which are at the same level and have little dispersion.
	
	
	\subsection{Reconstruction}
\begin{figure}[htbp]
    \centering
	\subfigure[y20 gw SET result, SNR = 228.6694]{
		\includegraphics[width=0.45\textwidth]{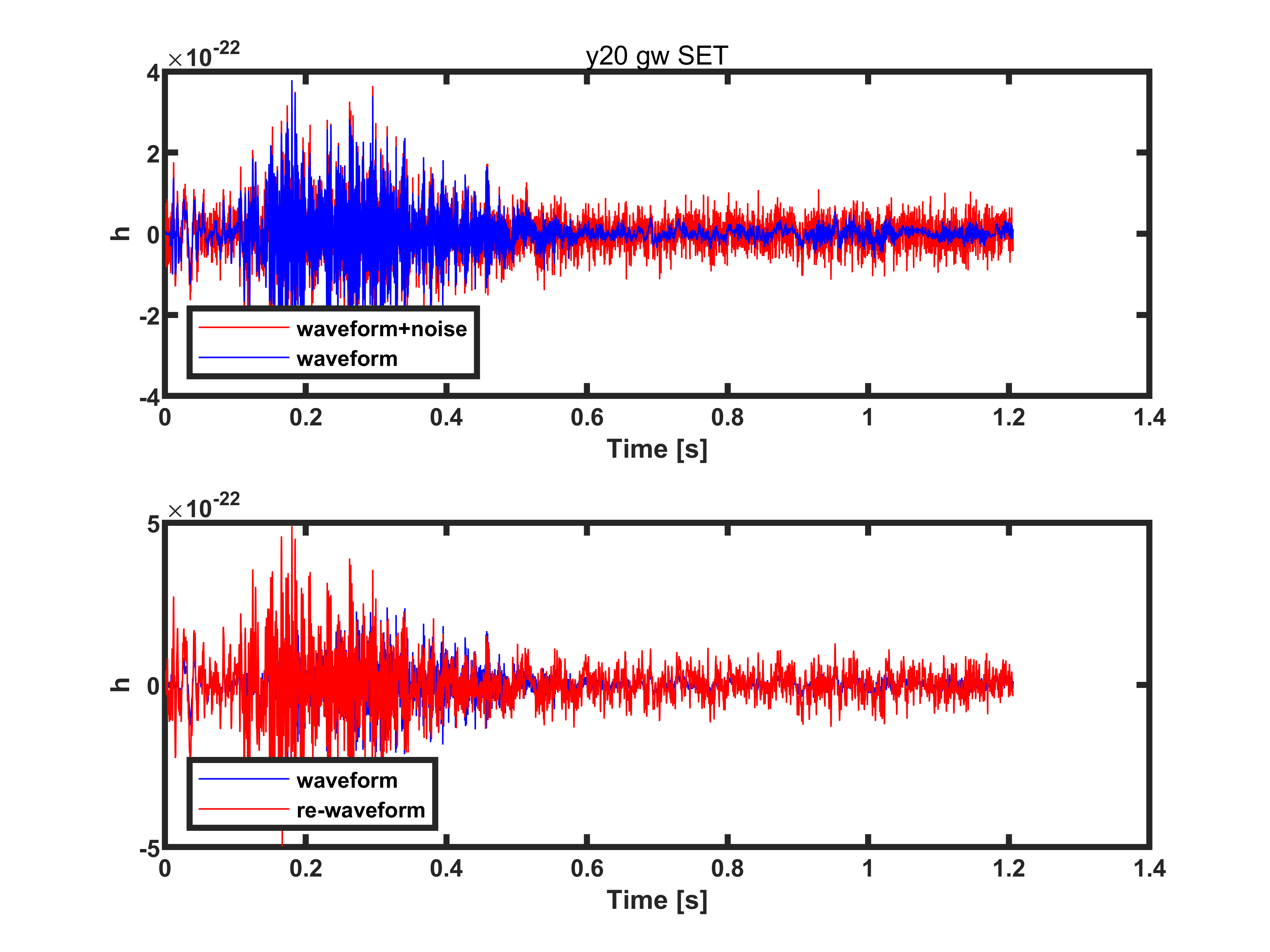}}
	\subfigure[y20 gw STFT result, SNR = 122.1948]{
		\includegraphics[width=0.45\textwidth]{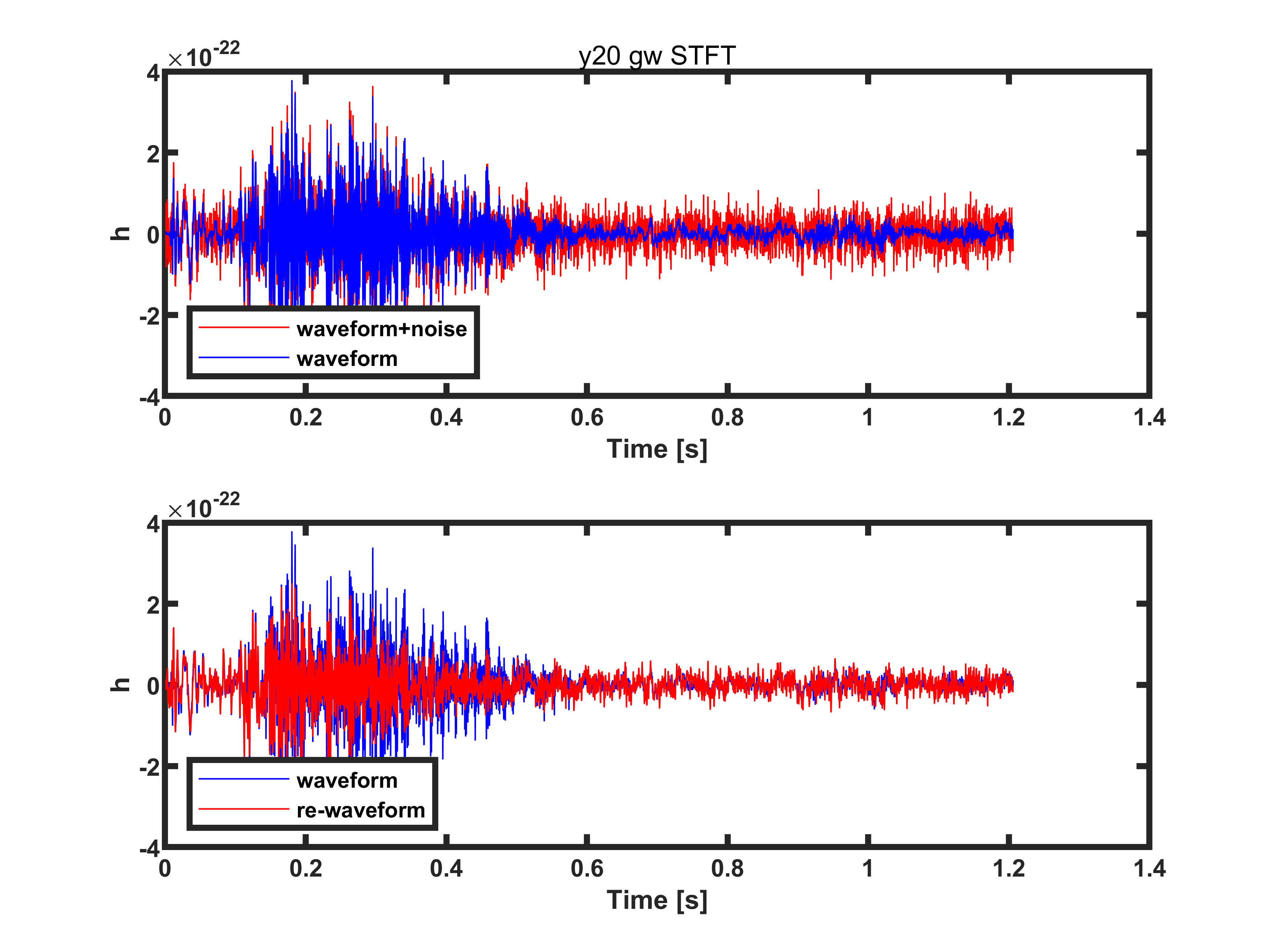}}
	\caption{These are the MSST, SET, and STFT results of our data. The x-axis represents time while y-axis represents the waveform. All the simulated signals are injected into white noise which is of the same average amplitude as the injected signal.}\label{exs}
    \end{figure}

\begin{figure}[htbp]
    	\centering
    	\subfigure[C15-3D GWsignal 0-438ms MSST result, 1 bin, SNR = 66.0792]{
    		\includegraphics[width=0.45\textwidth]{ft_sn_waveform_2019_3_26_re_IMSST.png}}
    	\subfigure[C15-3D GWsignal 0-438ms MSST result, 10 bins, SNR = 46.5656]{
    		\includegraphics[width=0.45\textwidth]{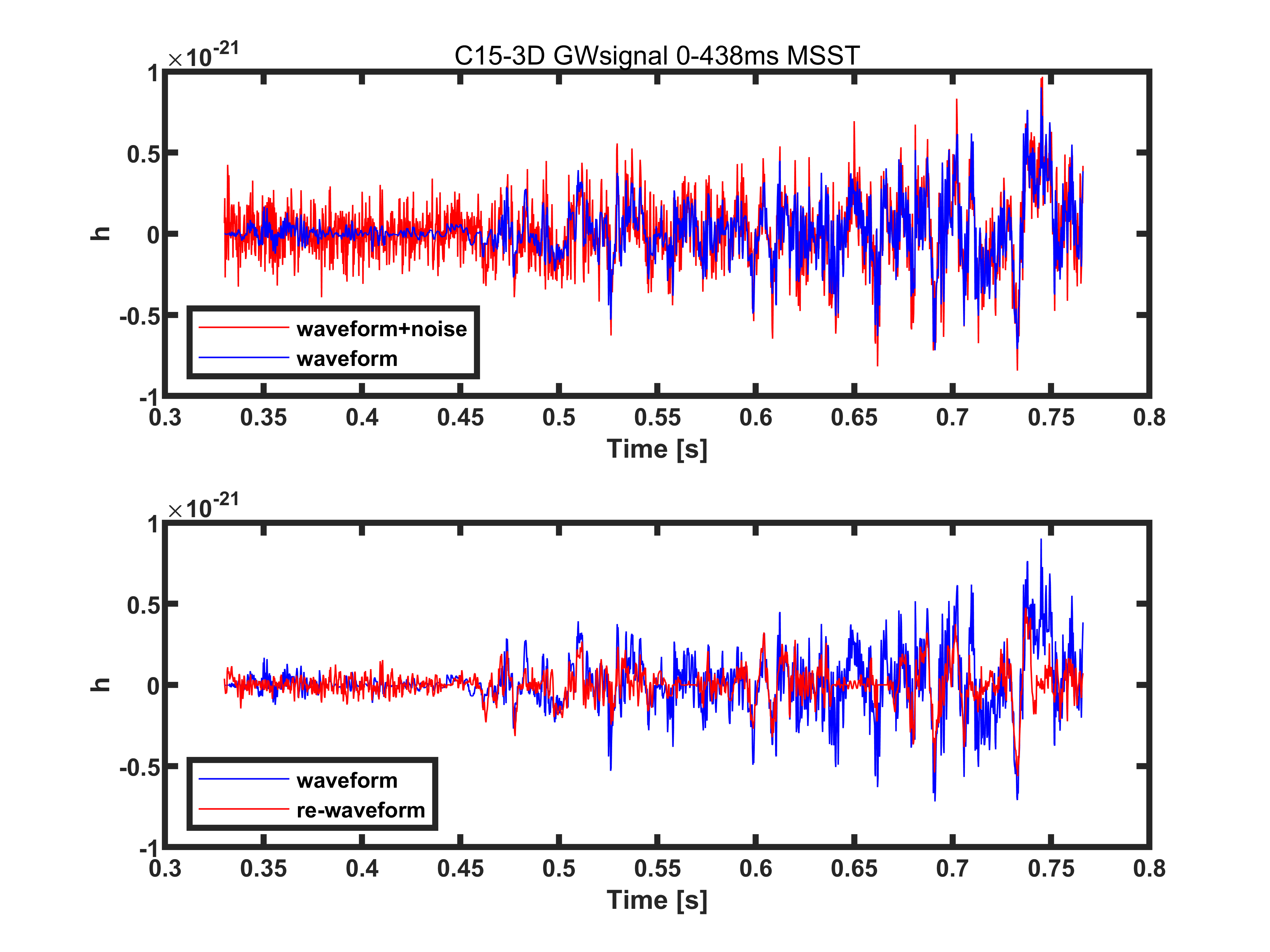}}
    	\caption{example to show the data bin effects}\label{bineffect}
    \end{figure}

	Results of simulated data mentioned in section II are shown in  figure \ref{exs}.    Obviously, the SET method performed the best. The average ratio between the SNR of SET results and the SNR of STFT results is 1.7421, while it is 1.9904 between the SET results and MSST results. The bad performance of the MSST method was mostly due to dividing the data into rather small bins. For the SET method and STFT method, dividing signals into bins and processing them separately has basically no effect on the reconstruction quality. However, we found that this step would lower the reconstruction quality of MSST results significantly. Fig \ref{bineffect}  show the effects of   dividing signals into bins for the performance of the MSST method, of signal "C15-3D GWsignal 0-438ms" which is injected into white noise with the same average amlitude as the signal.
	\begin{figure}[htbp]
		\centering
		\subfigure[C15-3D GWsignal 0-438ms MSST result, SNR = 68.3101]{
			\includegraphics[width=0.45\textwidth]{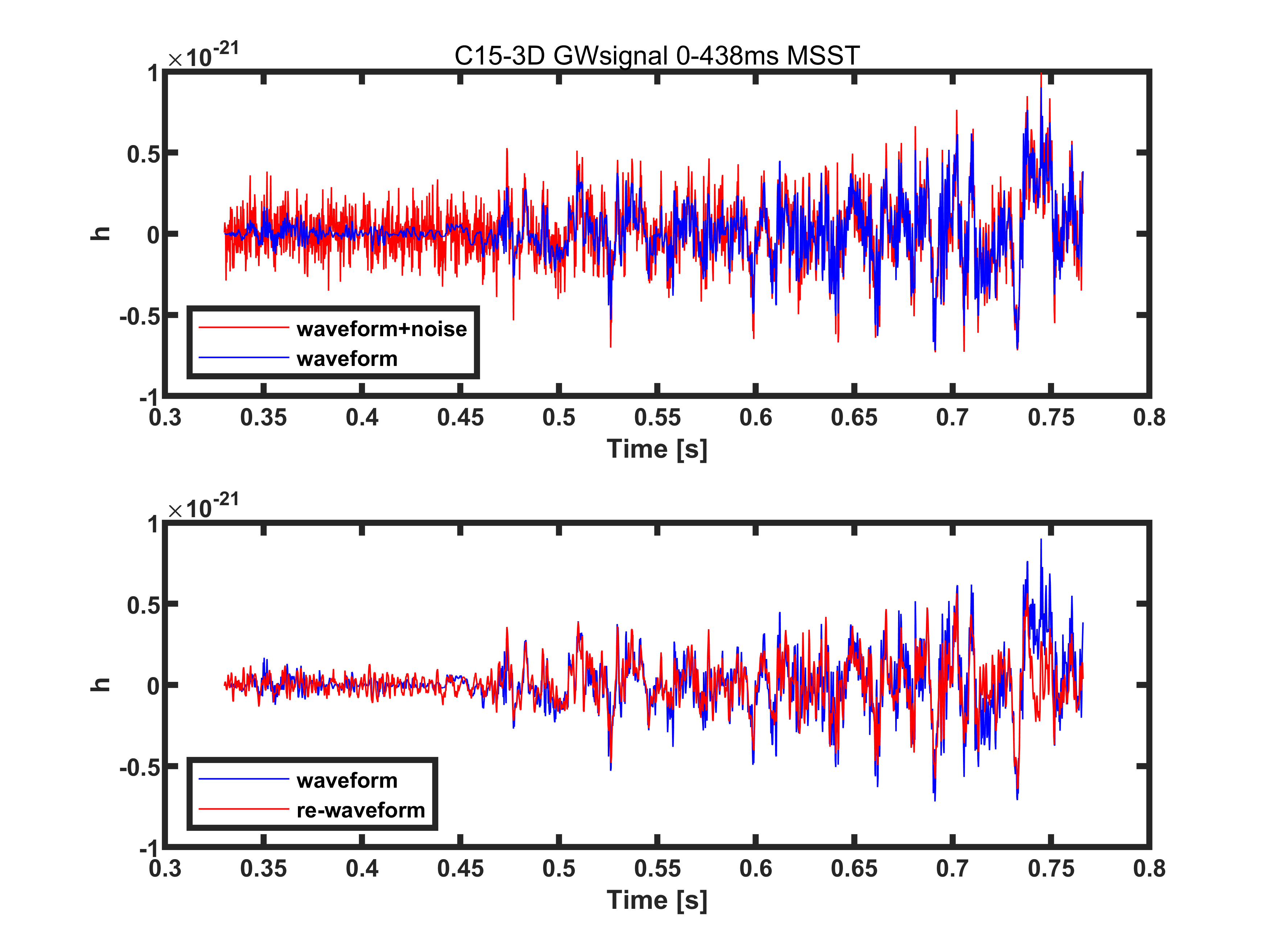}}
		\subfigure[C15-3D GWsignal 0-438ms SET result, SNR = 156.29665]{
			\includegraphics[width=0.45\textwidth]{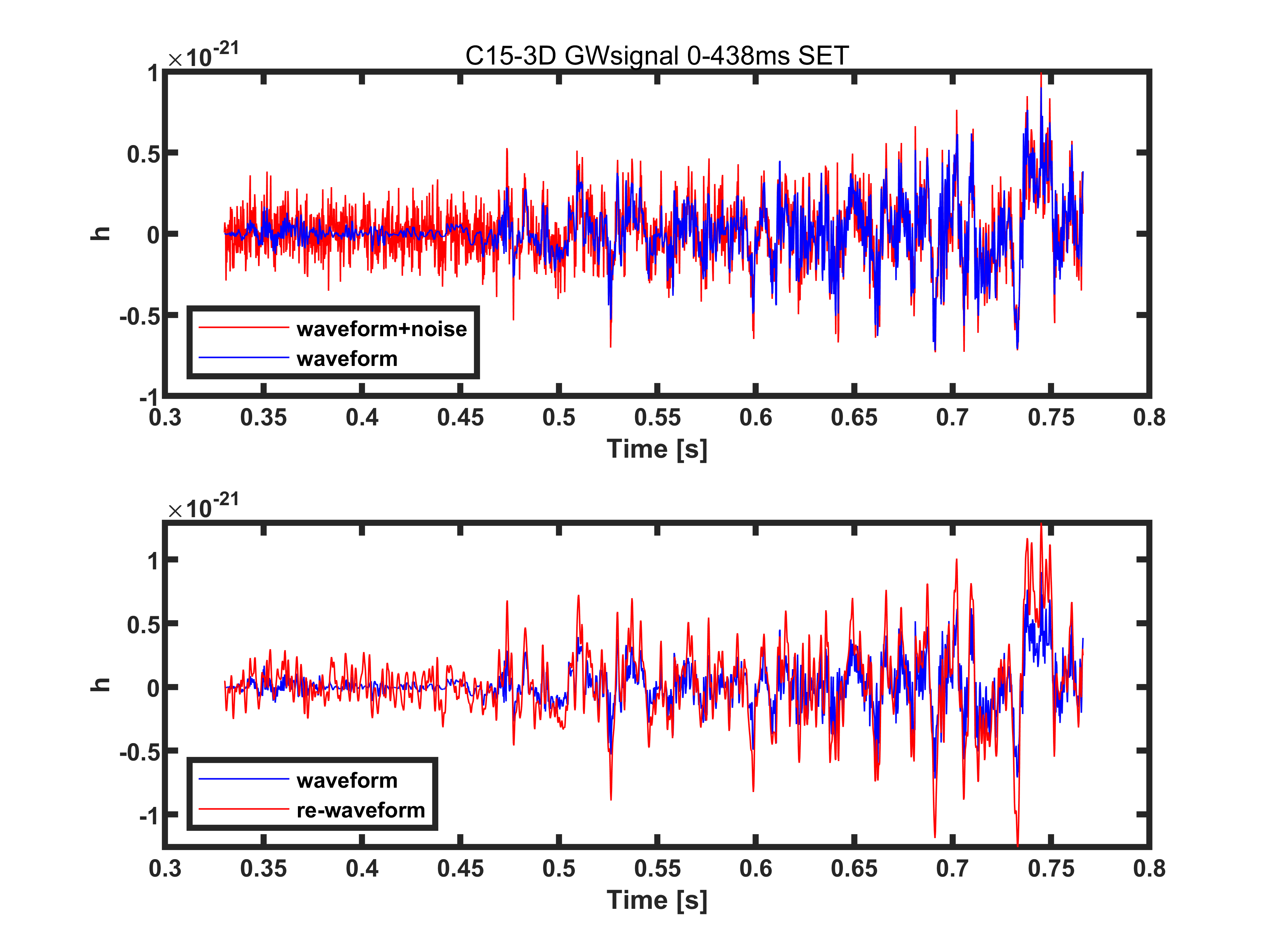}}
	\end{figure}

    \begin{figure}[htbp]
    	\centering
		\subfigure[C15-3D GWsignal 0-438ms STFT result, SNR = 90.9024]{
	    \includegraphics[width=0.45\textwidth]{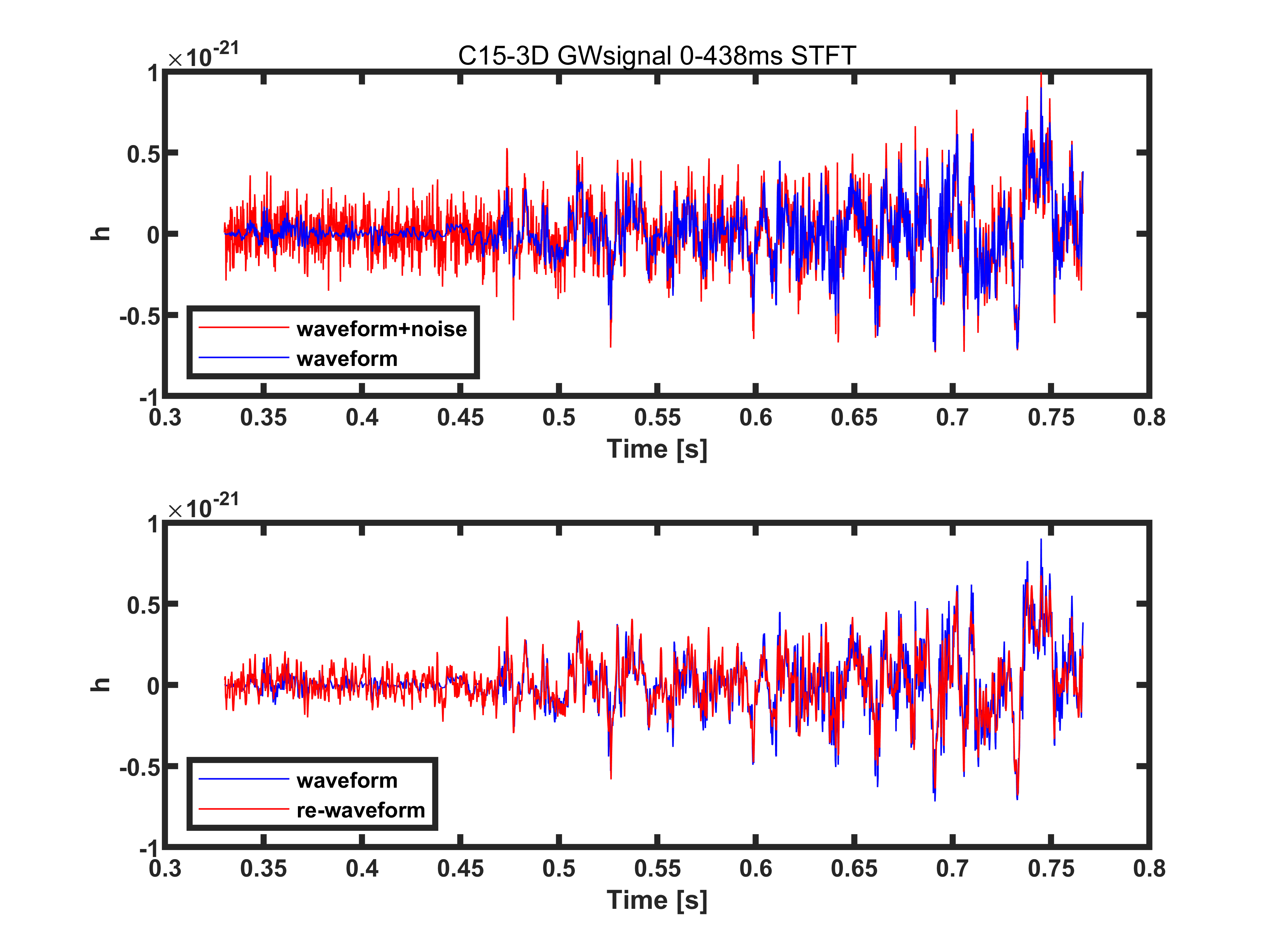}}
    \subfigure[m39gw MSST result, SNR = 198.3441]{
    	\includegraphics[width=0.45\textwidth]{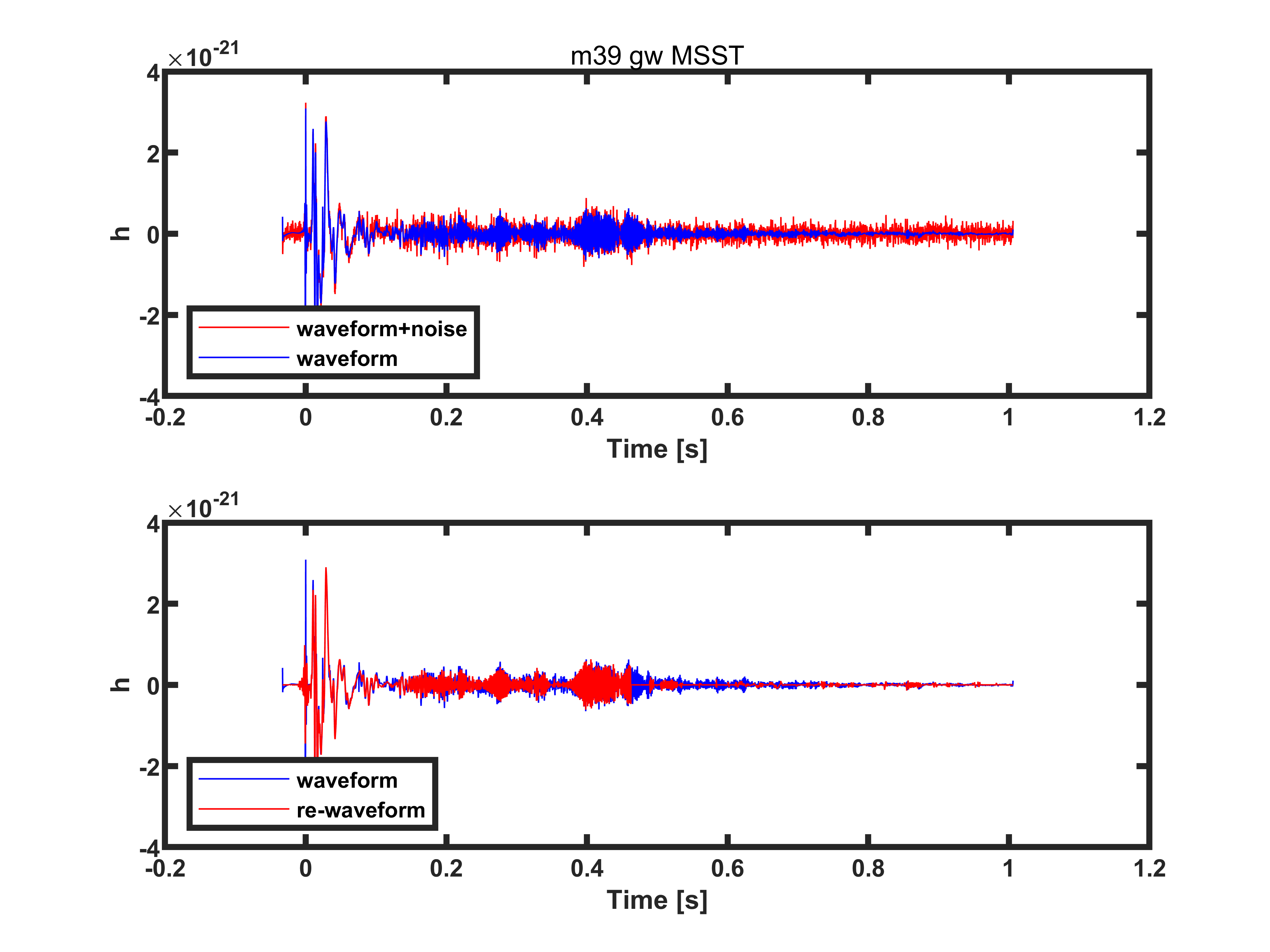}}
    \end{figure}

	\begin{figure}[htbp]
	\centering	
	\subfigure[m39gw SET result,SNR = 325.6475]{
		\includegraphics[width=0.45\textwidth]{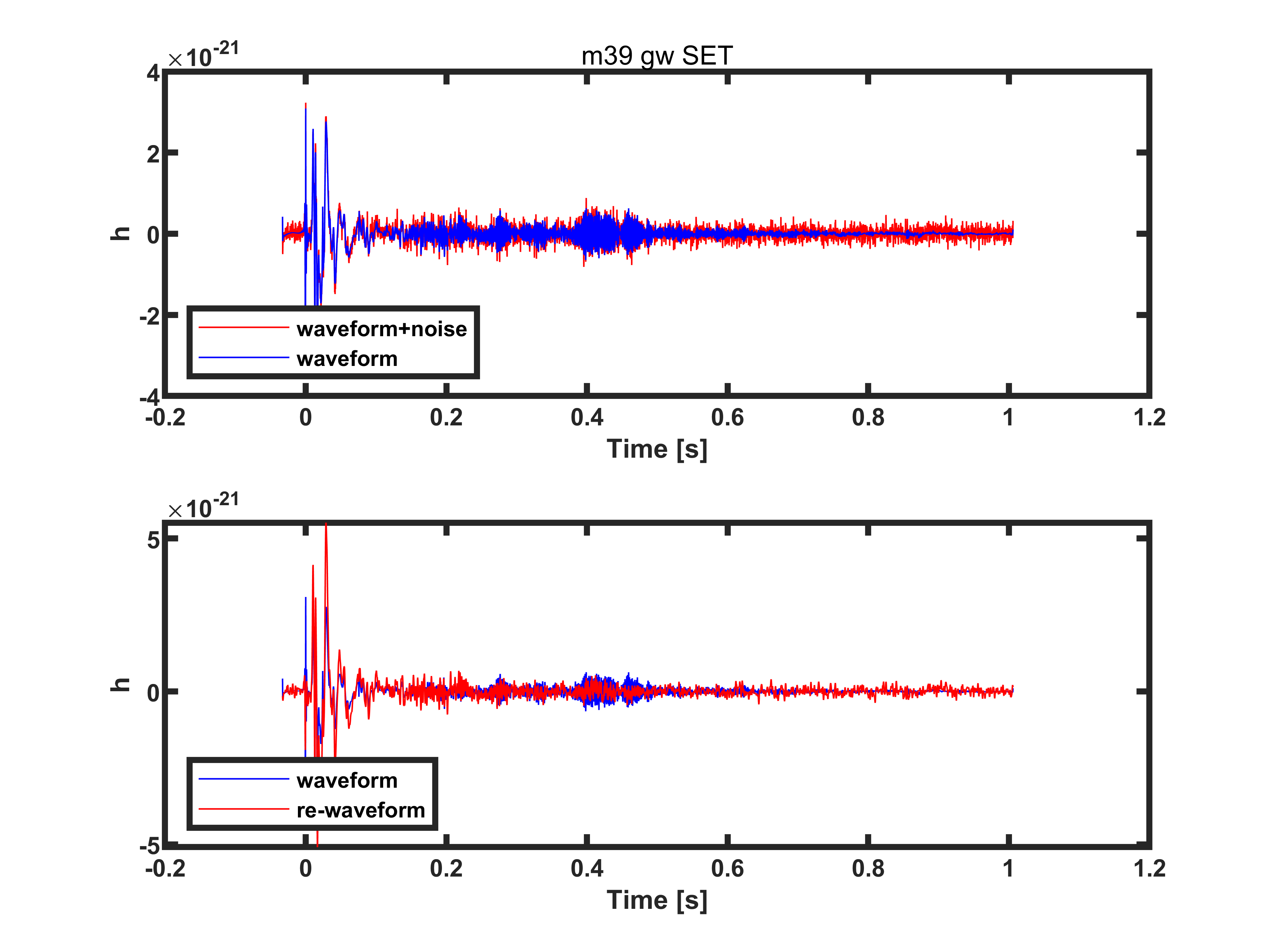}}
	\subfigure[m39gw STFT result, SNR = 161.1198]{
		\includegraphics[width=0.45\textwidth]{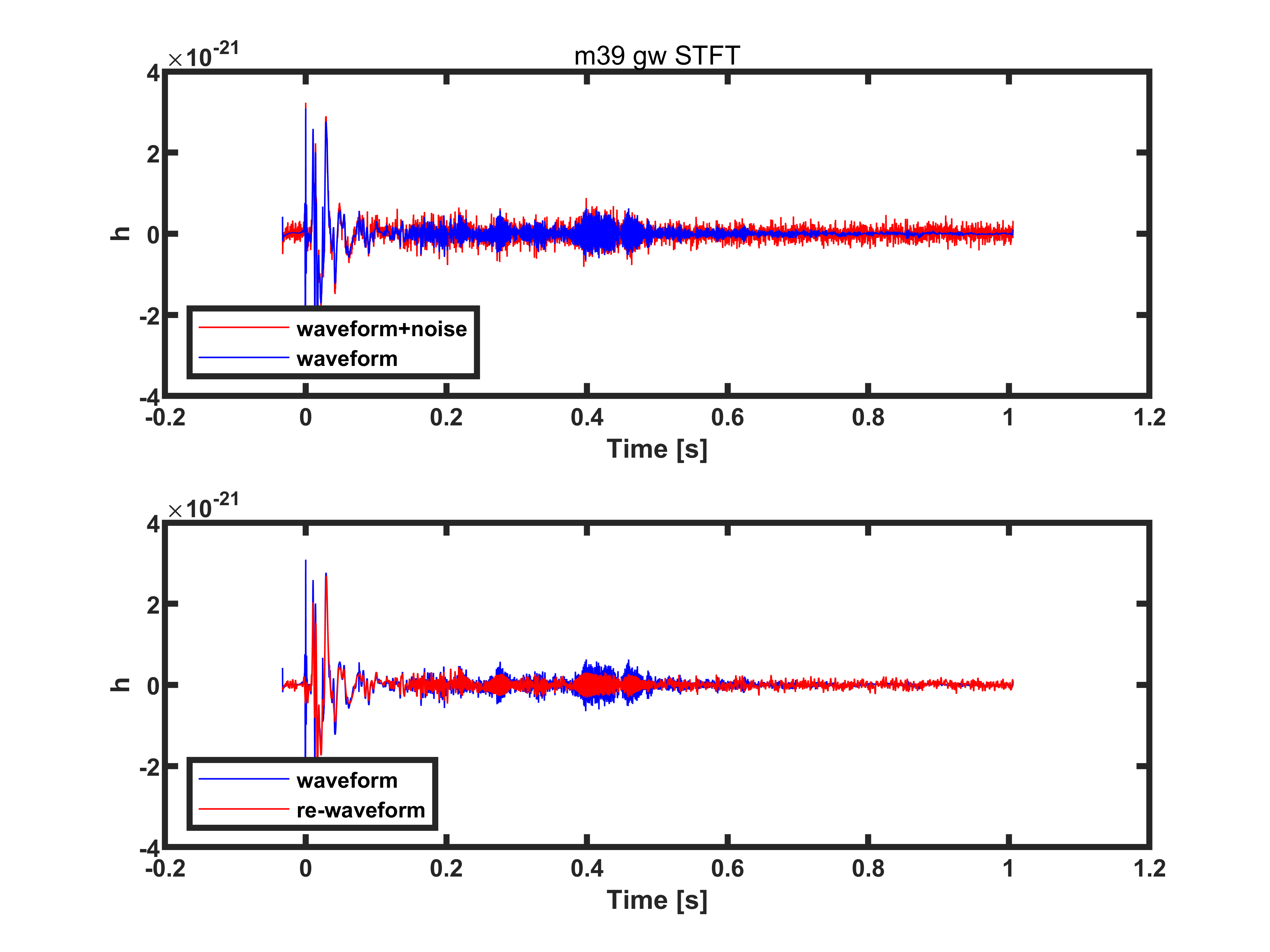}}
	\end{figure}

	\begin{figure}[htbp]
	\centering
	\subfigure[s3.5si gw MSST result, SNR = 111.1516]{
		\includegraphics[width=0.45\textwidth]{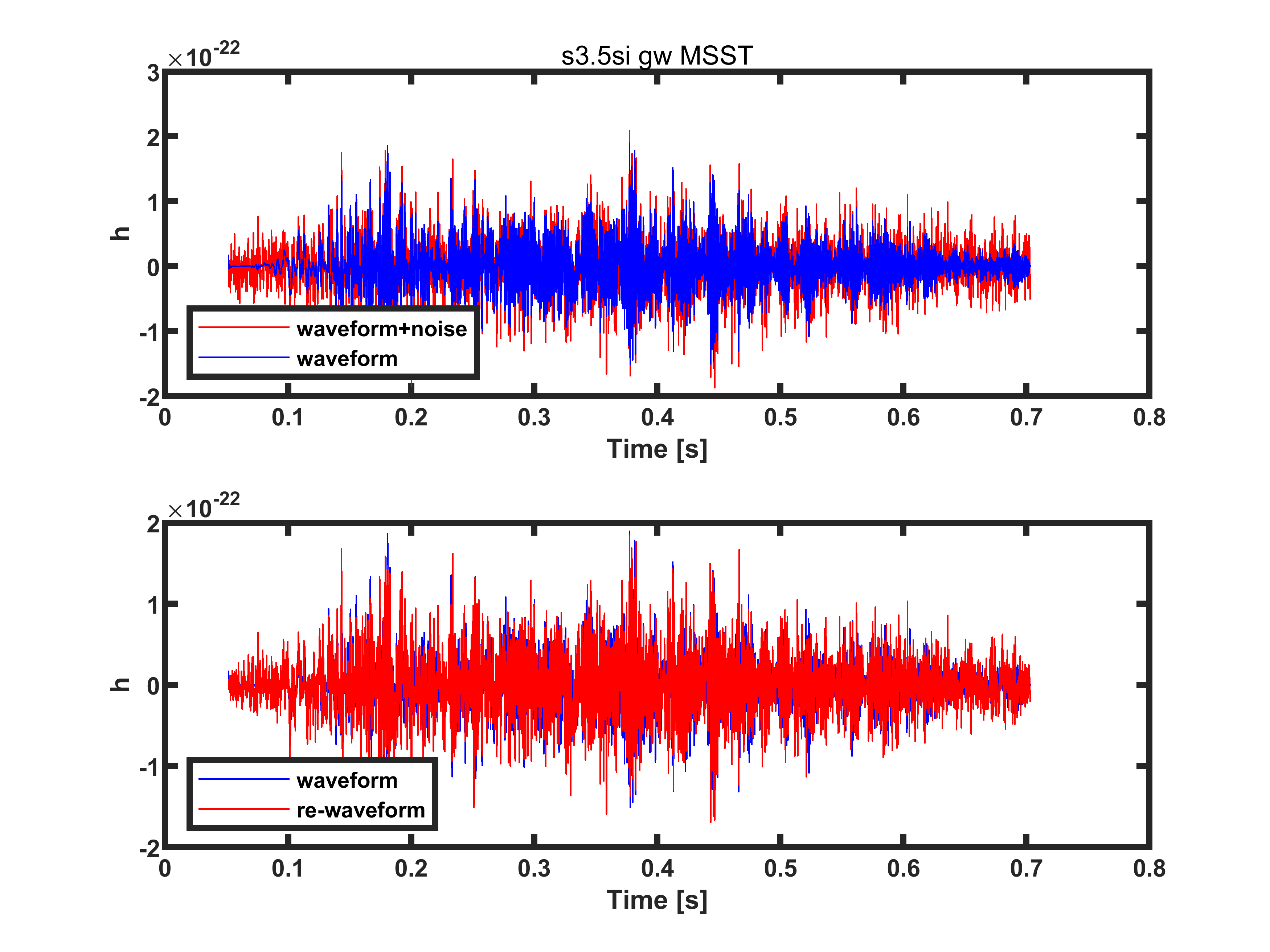}}
	\subfigure[s3.5si gw SET result, SNR = 209.0771]{
		\includegraphics[width=0.45\textwidth]{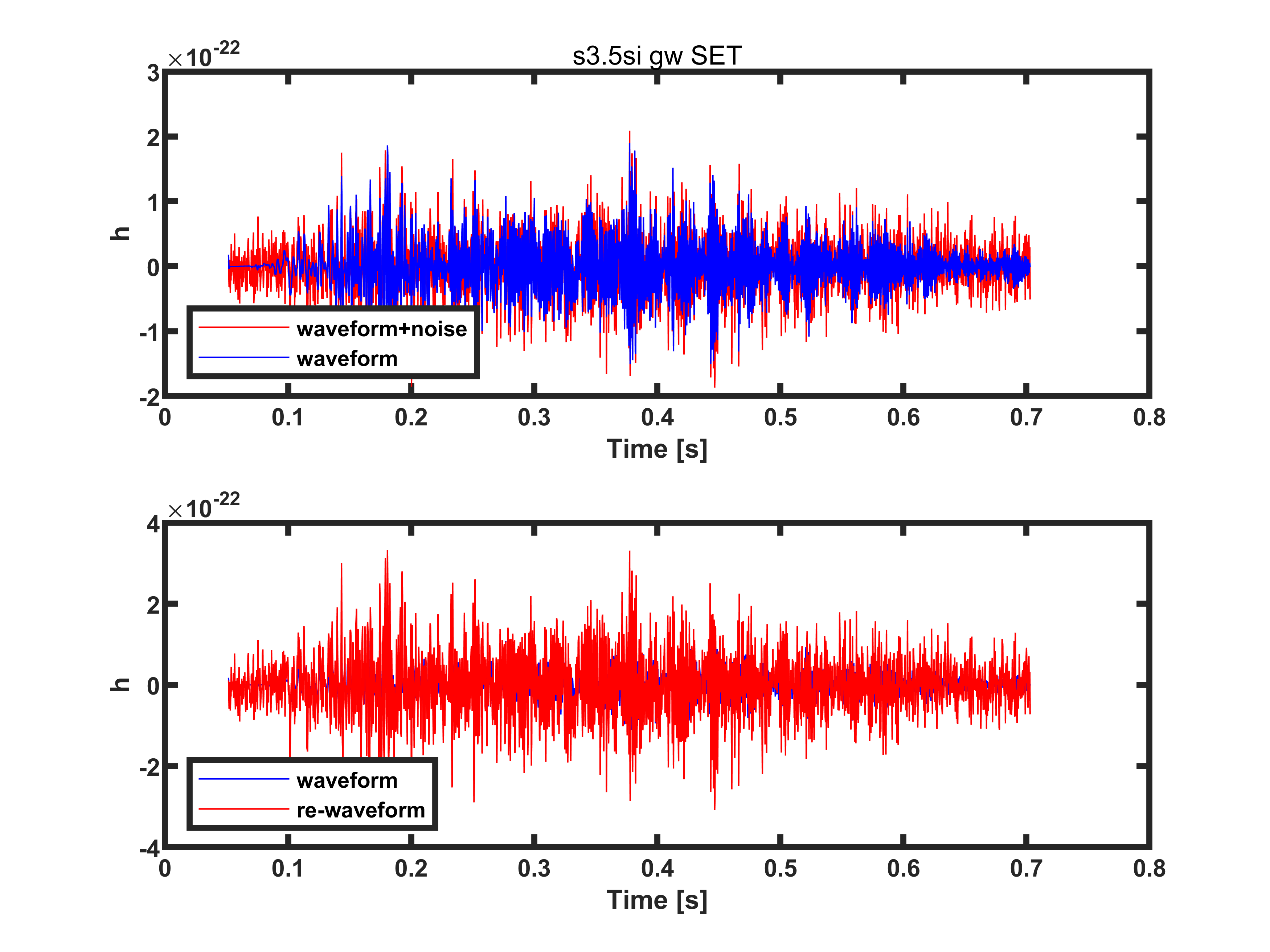}}
	\end{figure}

    \begin{figure}[htbp]
    \centering
    \subfigure[s3.5si gw STFT result, SNR = 124.8331]{
    	\includegraphics[width=0.45\textwidth]{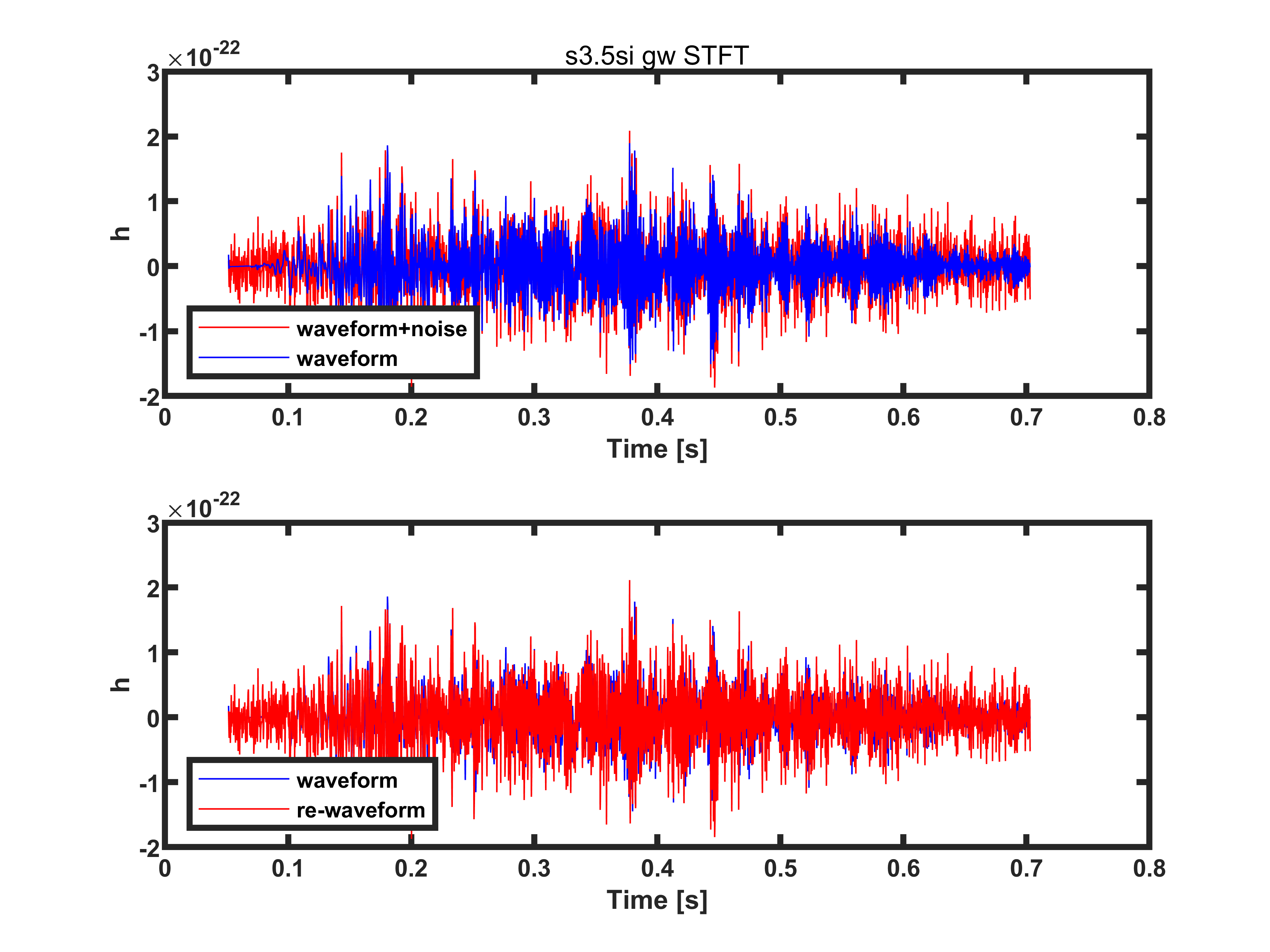}}
    \subfigure[s18 gw MSST result, SNR = 94.5154]{
    	\includegraphics[width=0.45\textwidth]{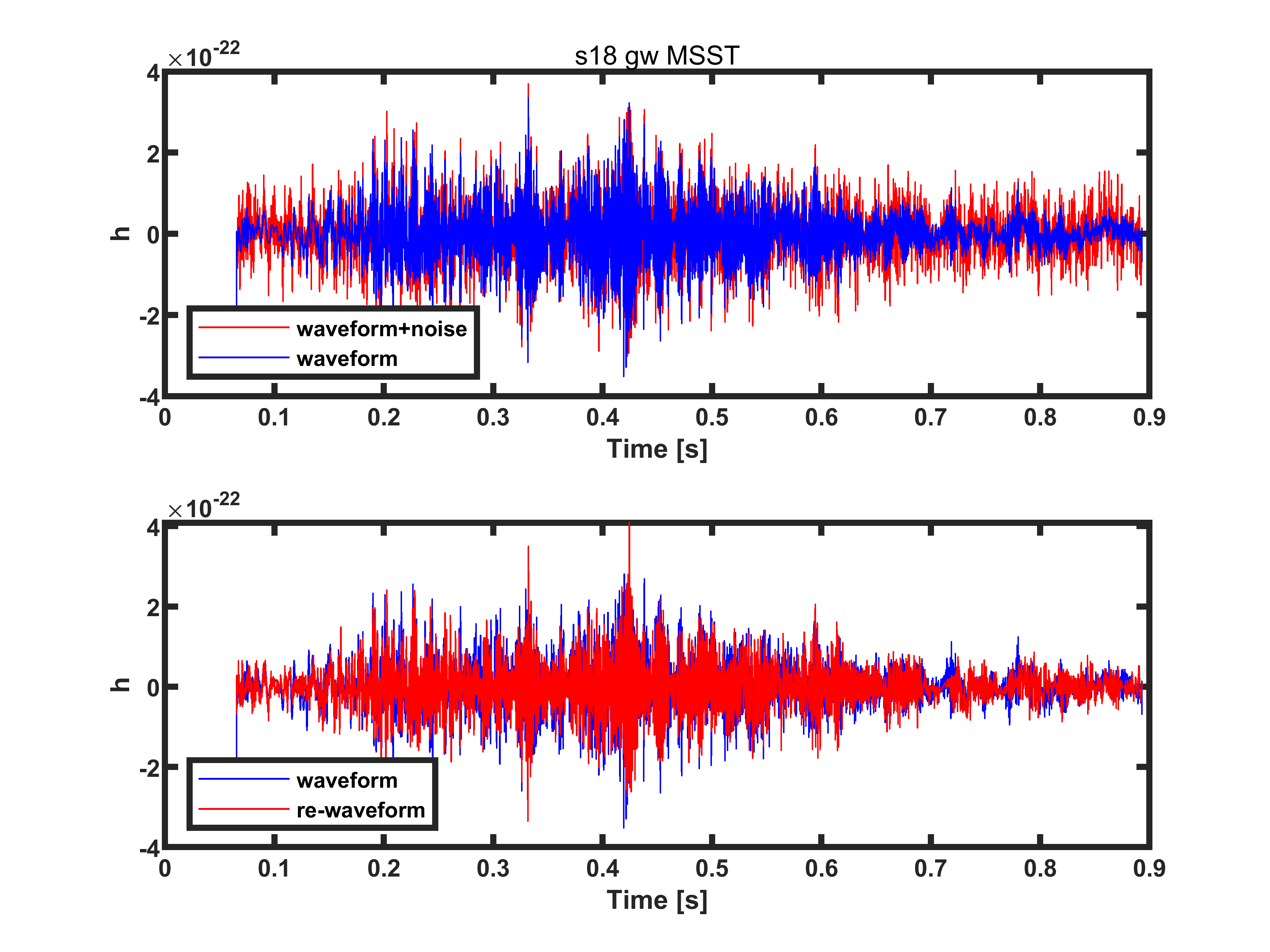}}
    \end{figure}

    \begin{figure}[htbp]
    \centering	
	\subfigure[s18 gw SET result, SNR = 243.1916]{
		\includegraphics[width=0.45\textwidth]{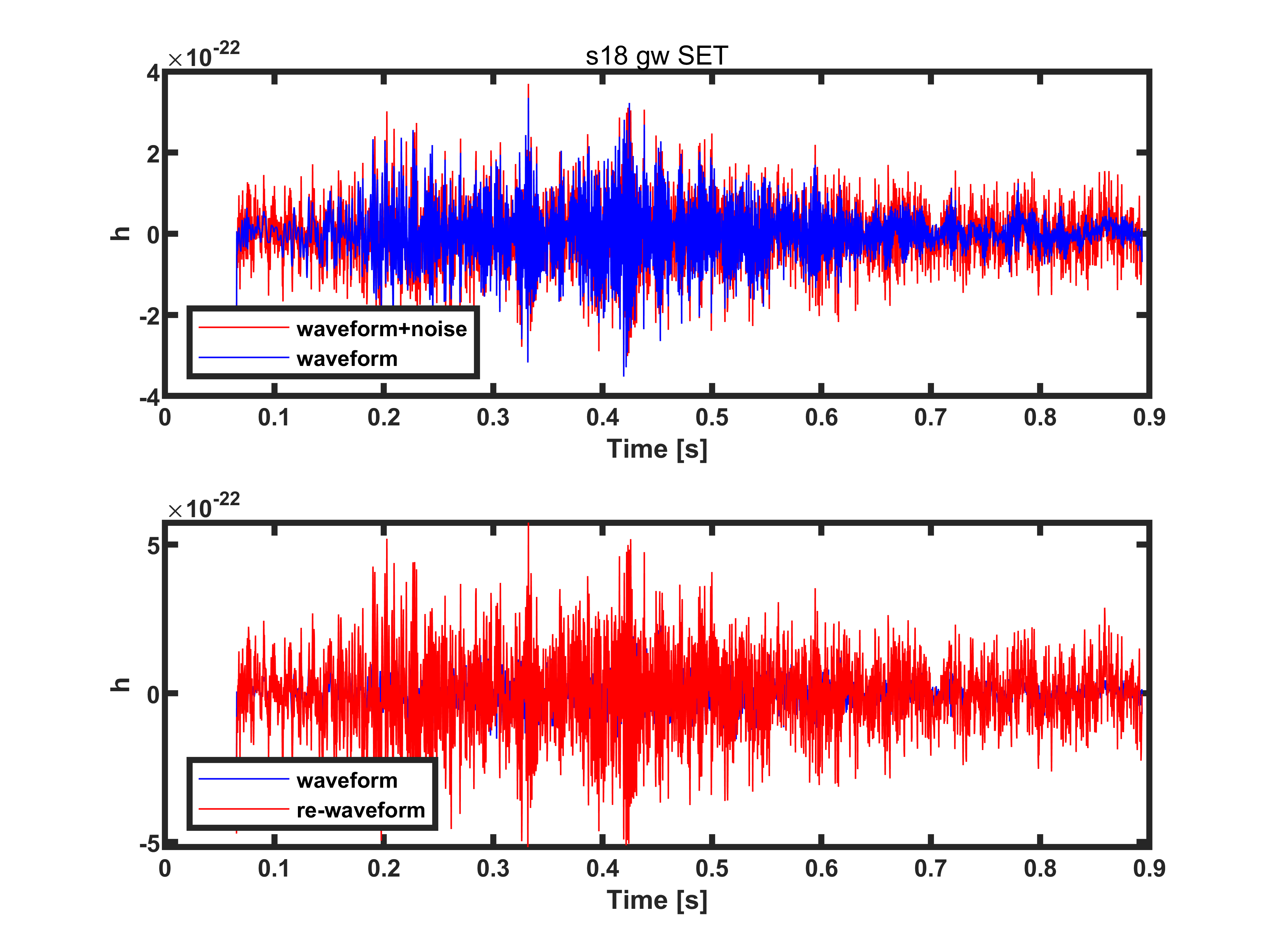}}
	\subfigure[s18 gw STFT result, SNR = 144.5712]{
		\includegraphics[width=0.45\textwidth]{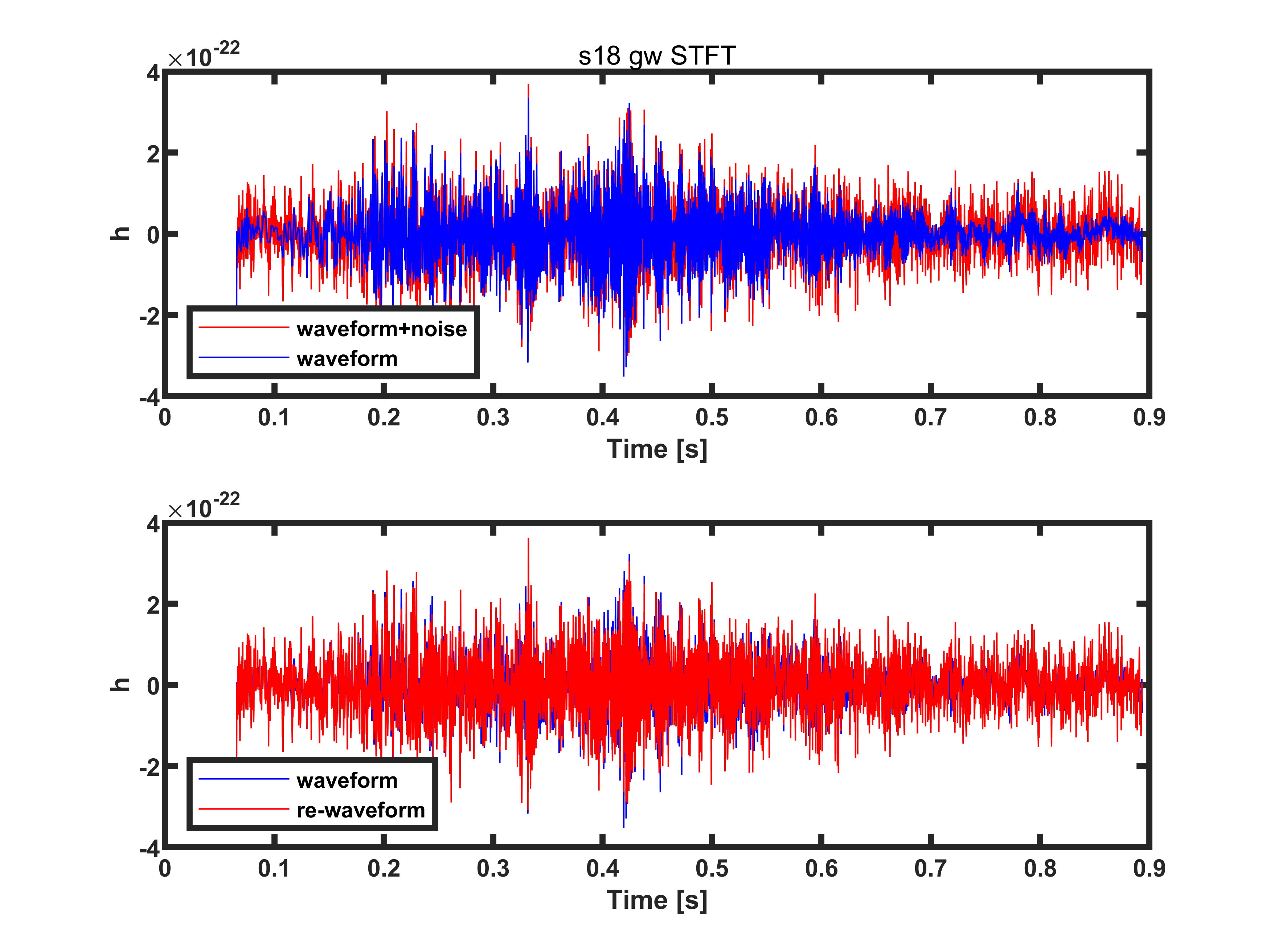}}
	\end{figure}

    \begin{figure}[htbp]
    \centering
	\subfigure[s18np gw MSST result, SNR = 113.1747]{
		\includegraphics[width=0.45\textwidth]{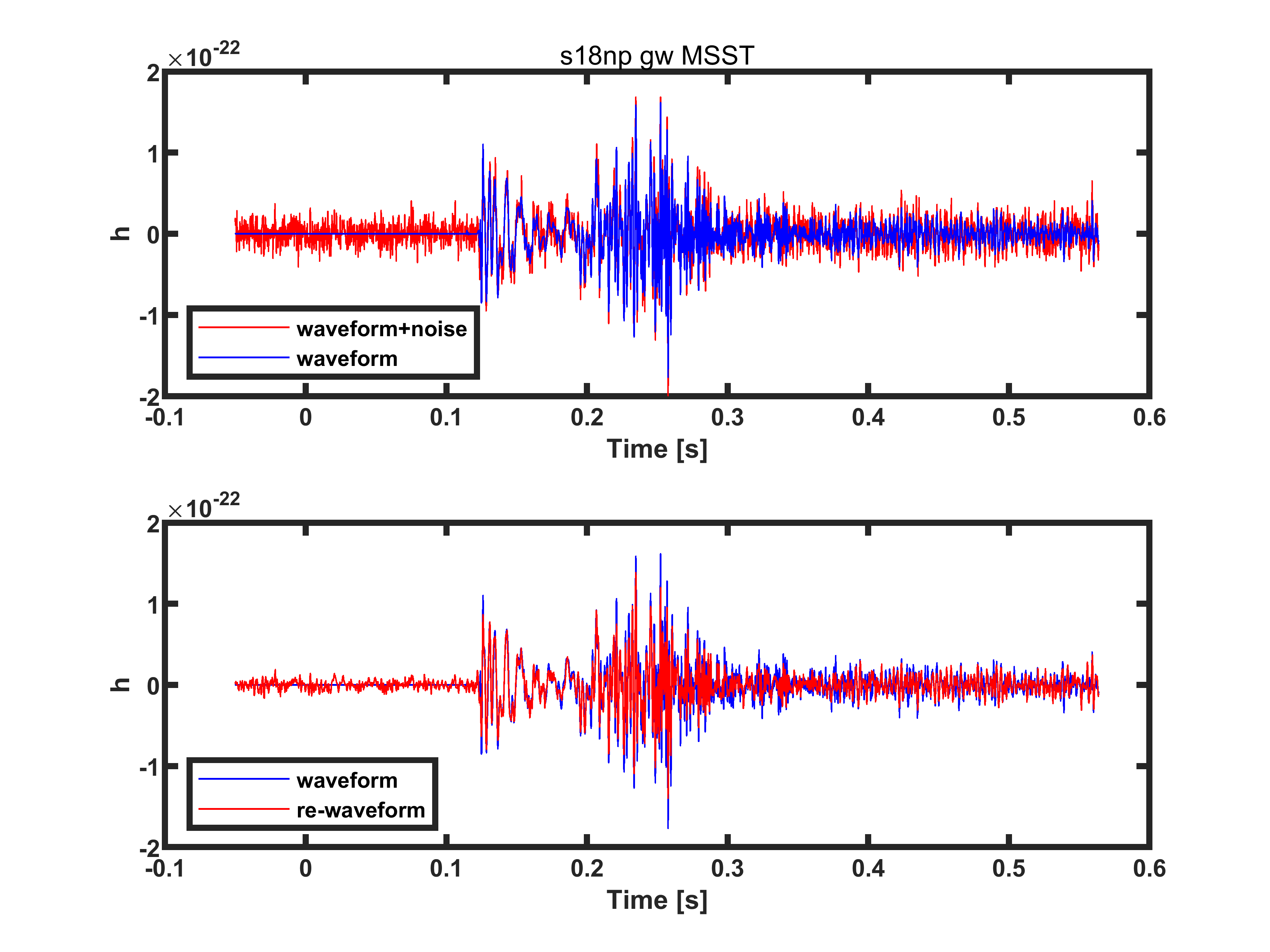}}
	\subfigure[s18np gw SET result, SNR = 173.9346]{
		\includegraphics[width=0.45\textwidth]{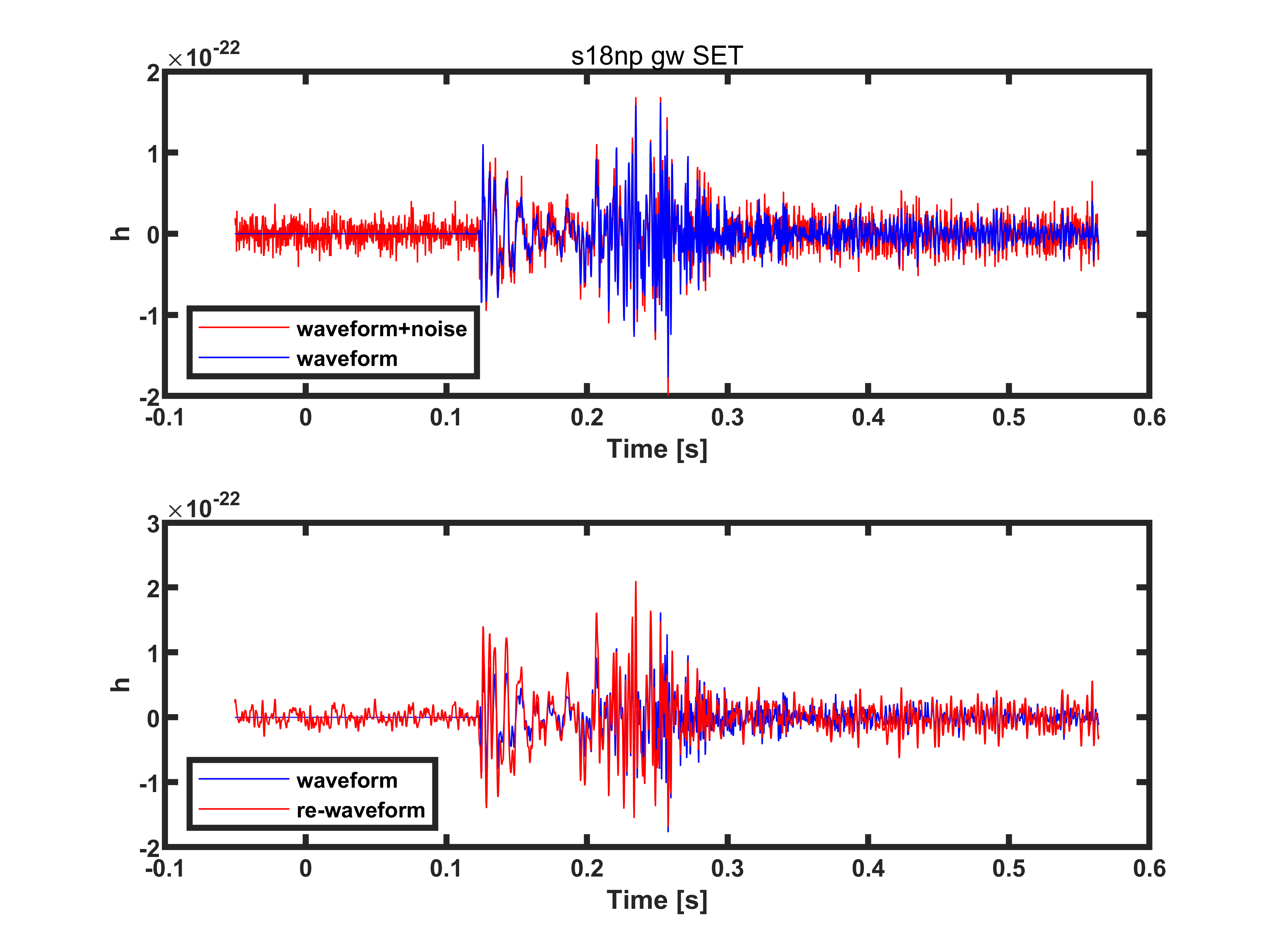}}
	\end{figure}

    \begin{figure}[htbp]
    	\centering
    	\subfigure[s18np gw STFT result, SNR = 116.7365]{
    		\includegraphics[width=0.45\textwidth]{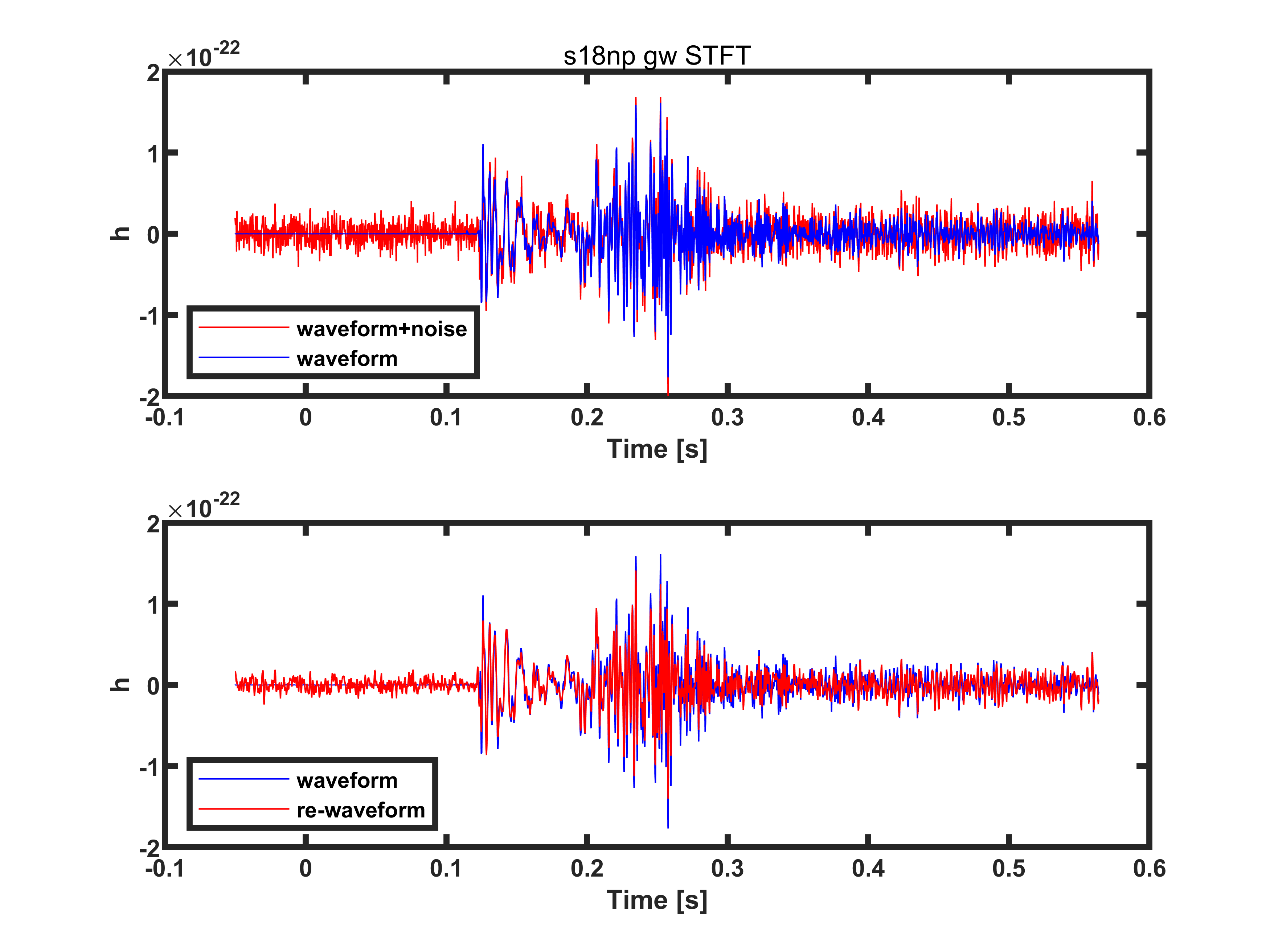}}
    		\subfigure[y20 gw MSST result, SNR = 113.0929]{
    		\includegraphics[width=0.45\textwidth]{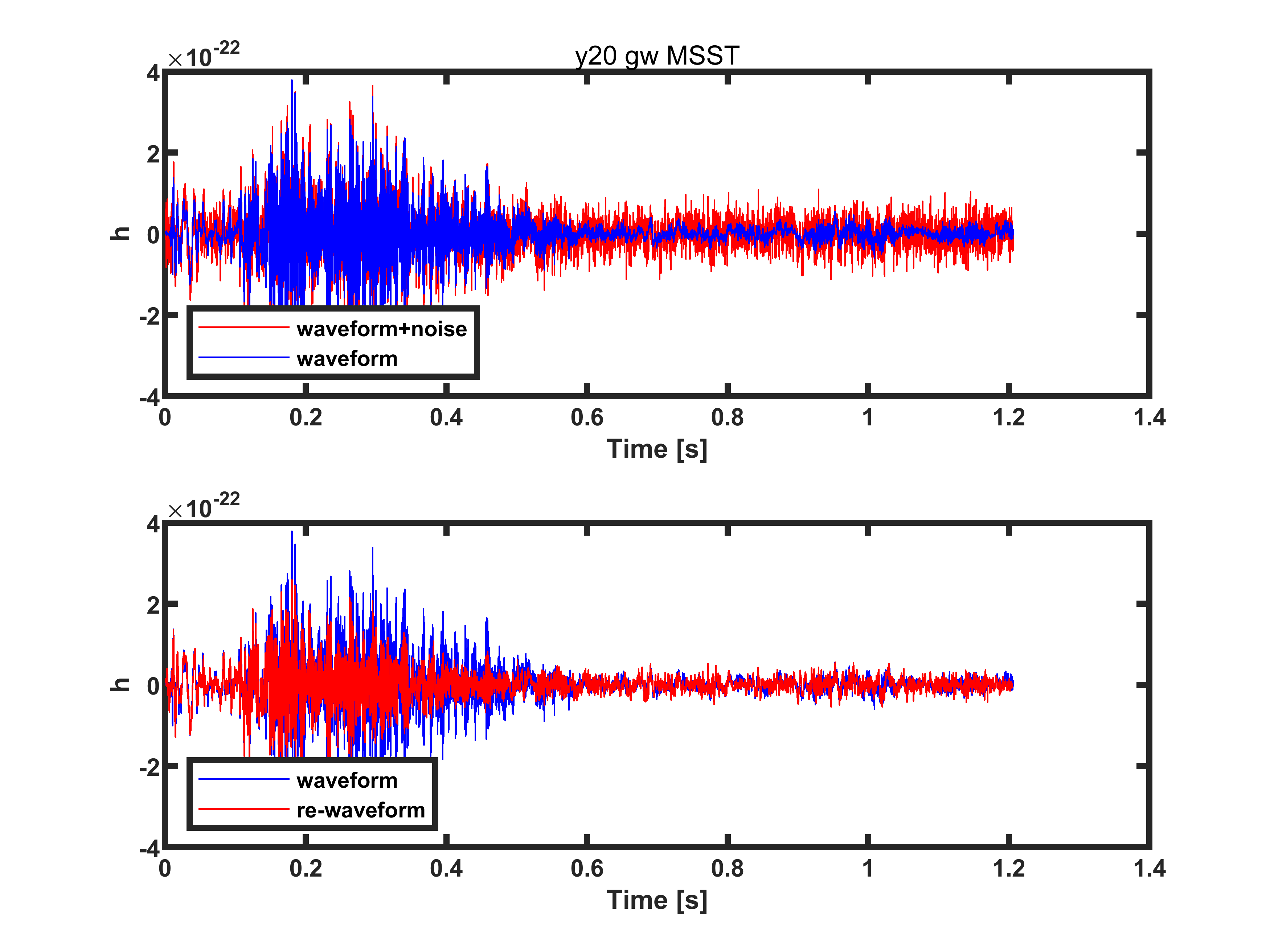}}

    \end{figure}
Another notable fact is the short window length of the SET method needed. Reconstruction of the SET method under the same window length with other methods, 250 in this paper, will leak most of the energy in the high-frequency band, leaving only low-frequency mode, which can be useful to detect the low-frequency mode under high-frequency noise. But when the signals are only in the high-frequency band without any low-frequency mode, or if one needs all the information in all frequency band, the window length must be very small, we took 10 in this paper, to counteract this effect. An example is shown in Fig \ref{wineffect}:
 \begin{figure}[htbp]
    	\centering
    	\subfigure[C15-3D GWsignal 0-438ms SET result, window length = 10]{
    		\includegraphics[width=0.45\textwidth]{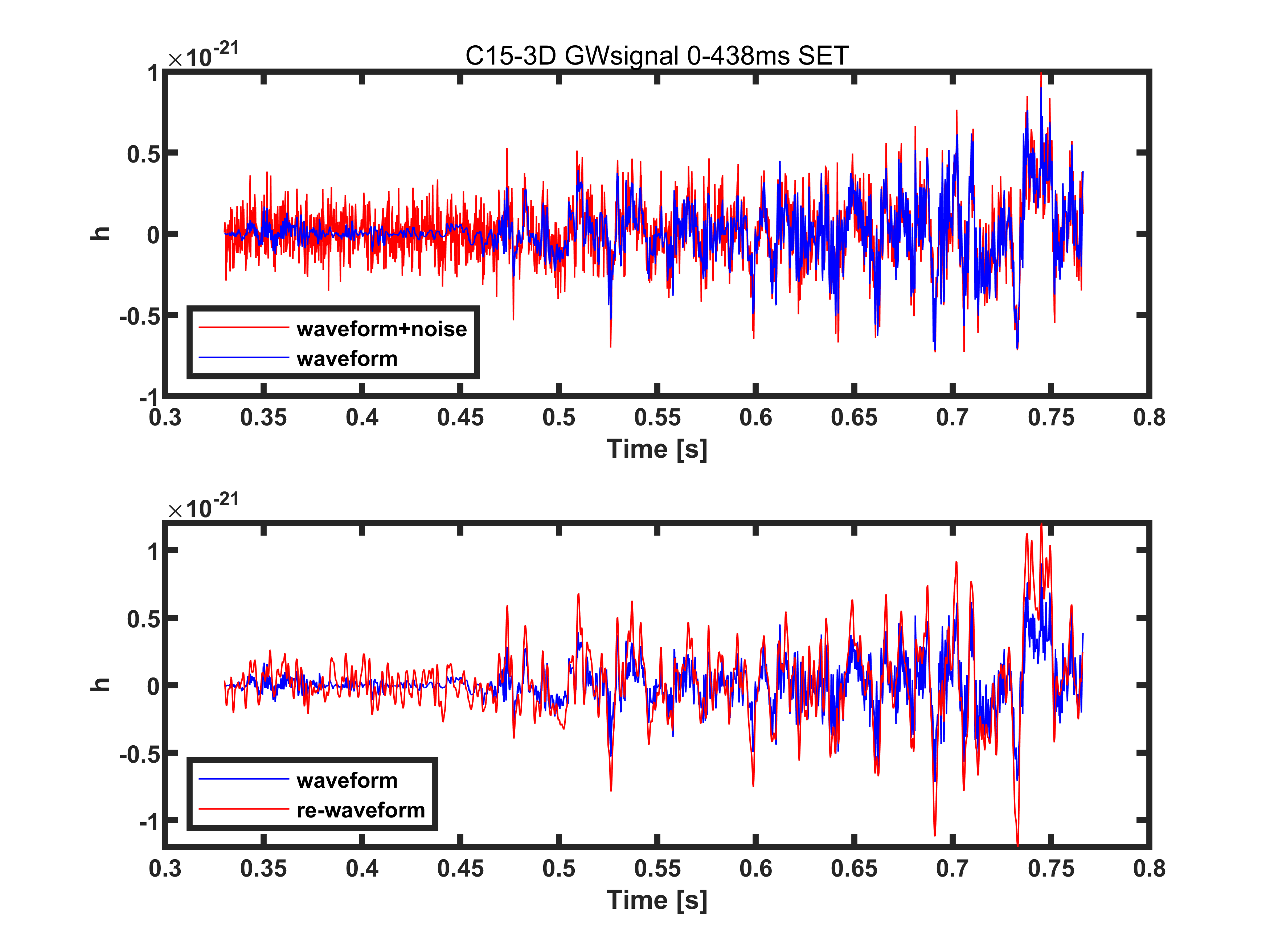}}
    	\subfigure[C15-3D GWsignal 0-438ms SET result, window length = 100]{
    		\includegraphics[width=0.45\textwidth]{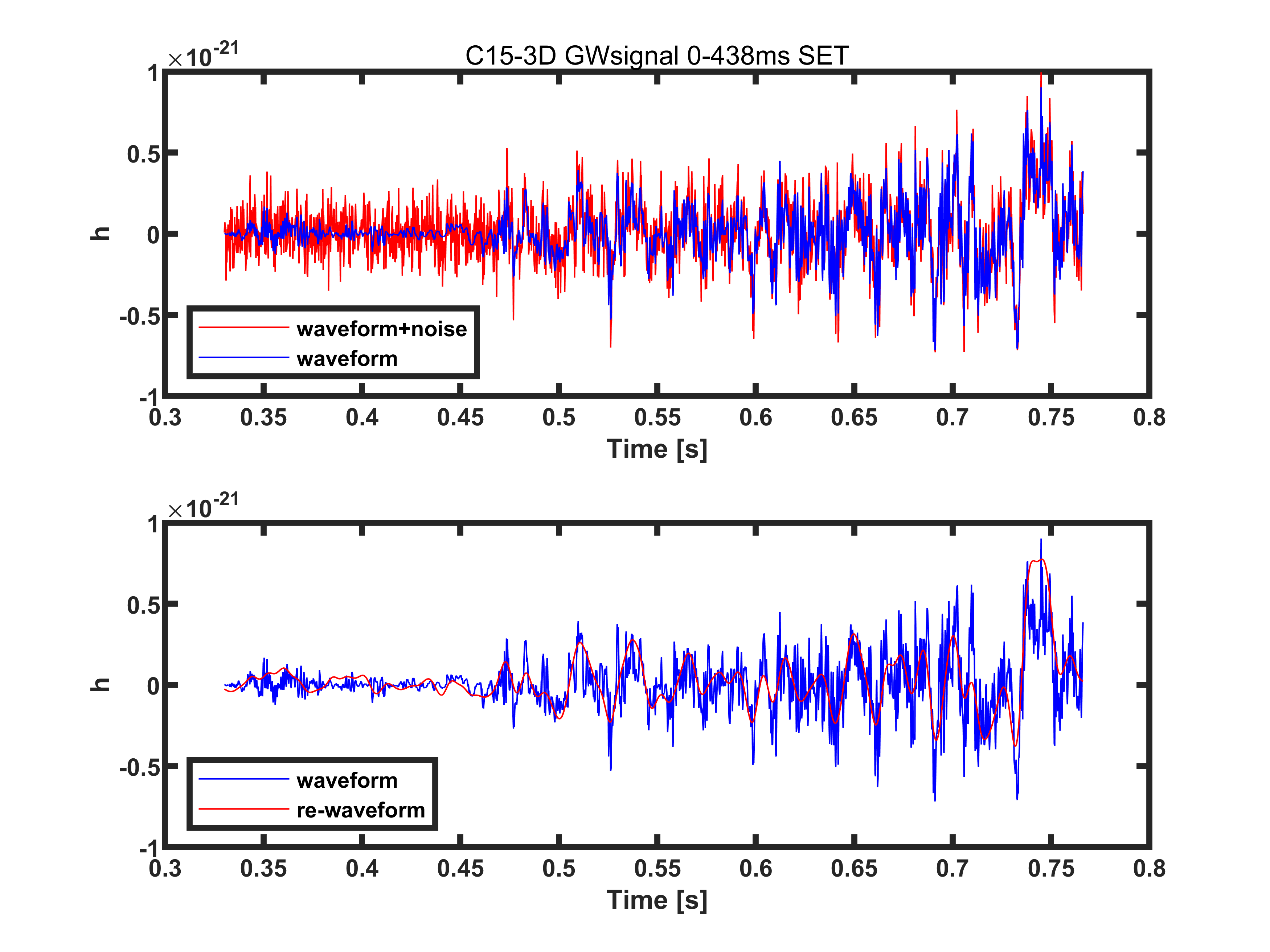}}
    	\subfigure[C15-3D GWsignal 0-438ms MSST result, window length = 500]{
    		\includegraphics[width=0.45\textwidth]{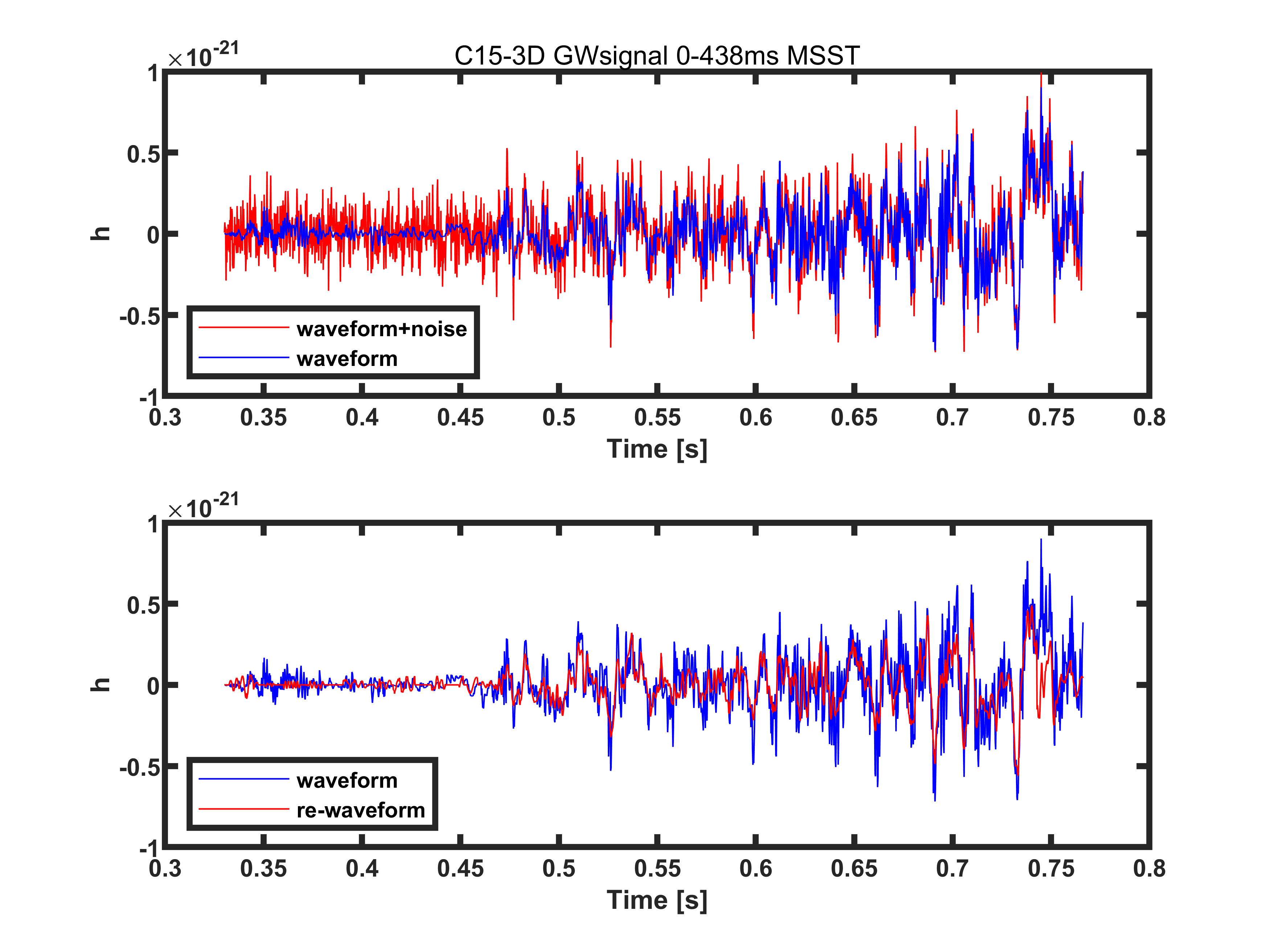}}
    	\subfigure[C15-3D GWsignal 0-438ms STFT result, window length = 500]{
    		\includegraphics[width=0.45\textwidth]{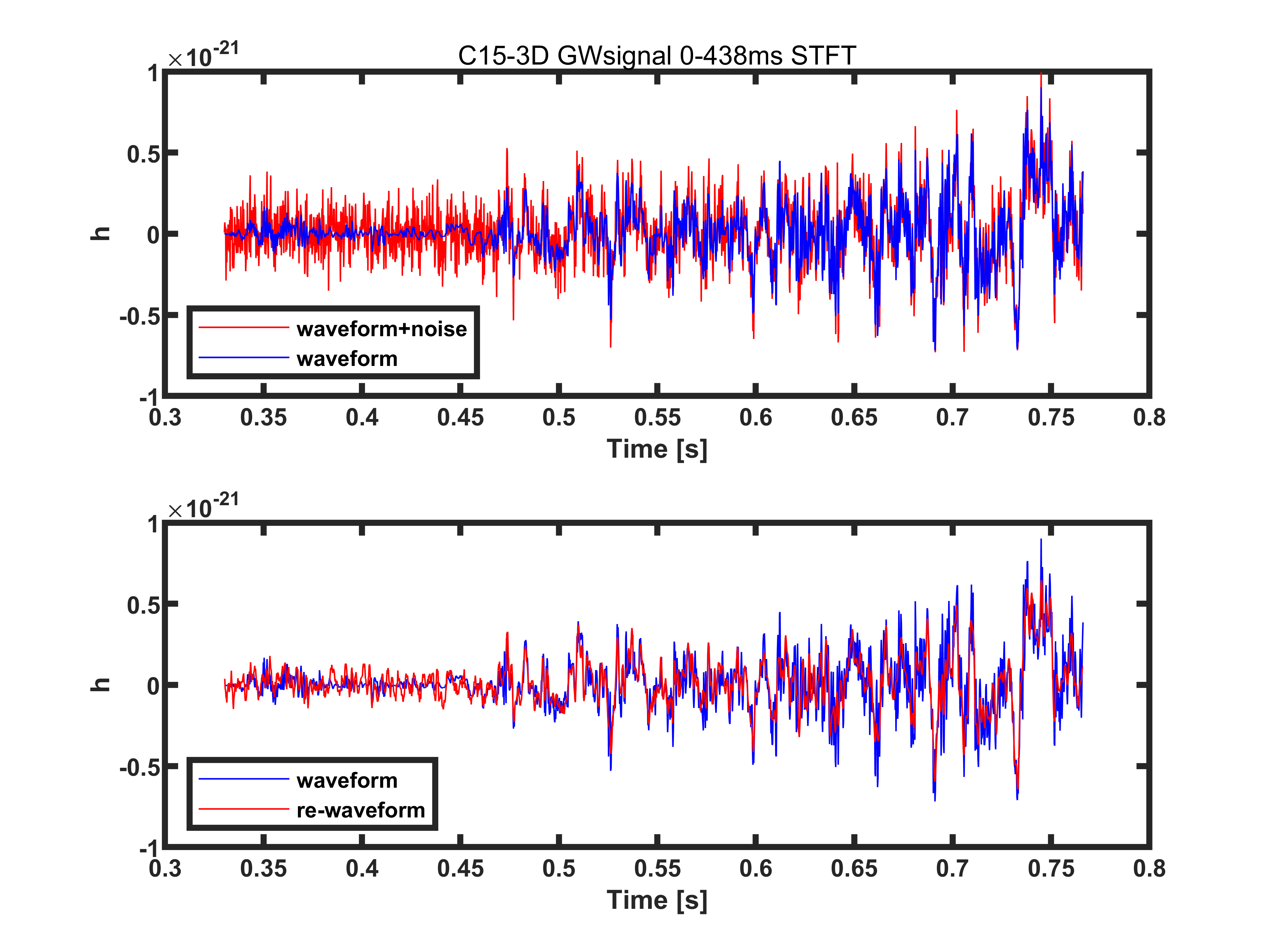}}
    		\caption{example to show the window length effects}\label{wineffect}
    \end{figure}
   From these results we expect that the SET result using a longer window can extract the low-frequency mode very efficiently, while other methods can not achieve this by only modifying the window length. However more tests are still needed.
	\subsection{Different  noise level tests}
\begin{figure}[htbp]
		\centering
		\subfigure[MSST result, double the noise, SNR = 31.7741]{
			\includegraphics[width=0.45\textwidth]{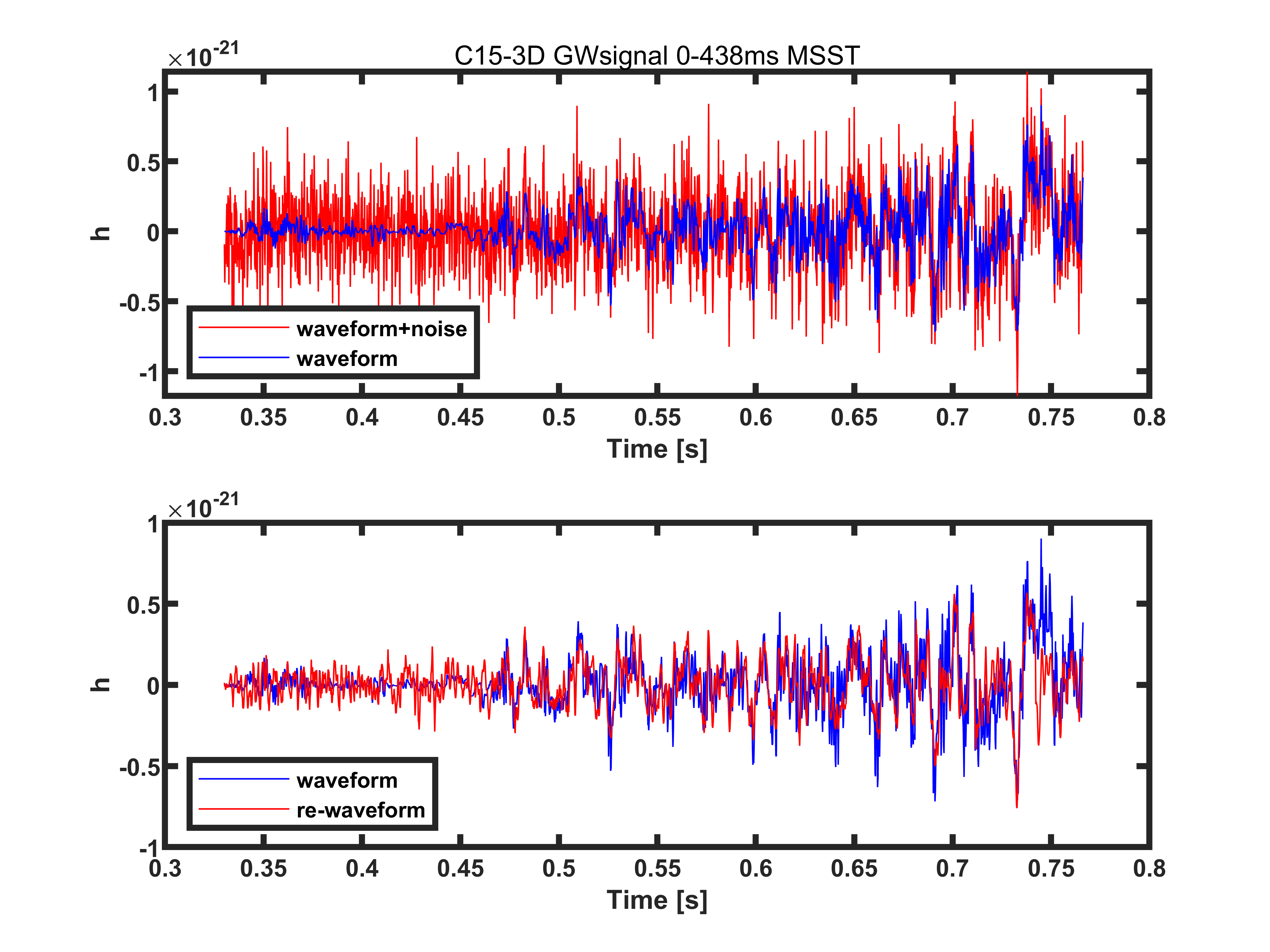}}
		\subfigure[SET result, double the noise, SNR = 77.7253]{
			\includegraphics[width=0.45\textwidth]{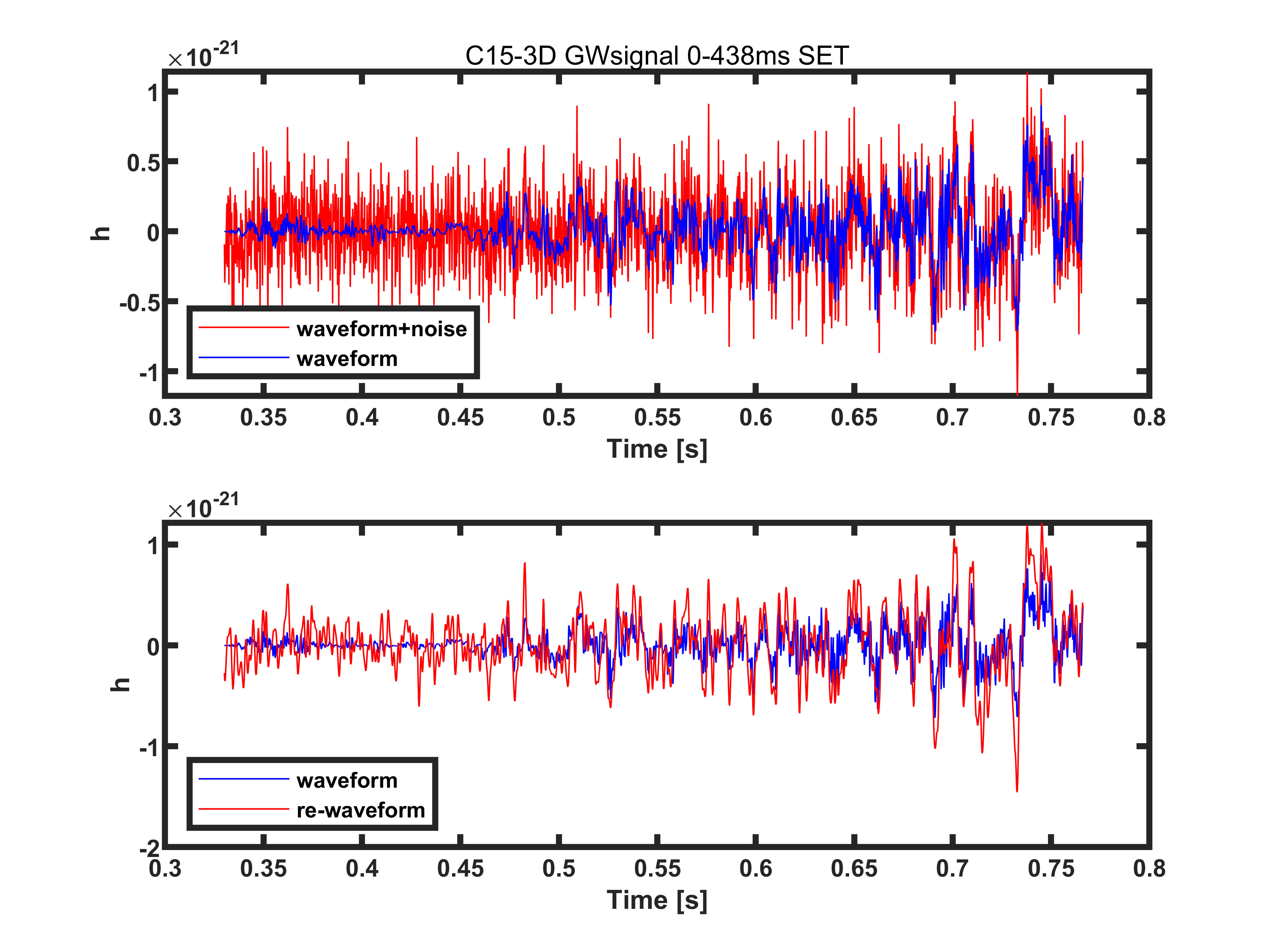}}
		\subfigure[STFT result, double the noise, SNR = 40.7937]{
			\includegraphics[width=0.45\textwidth]{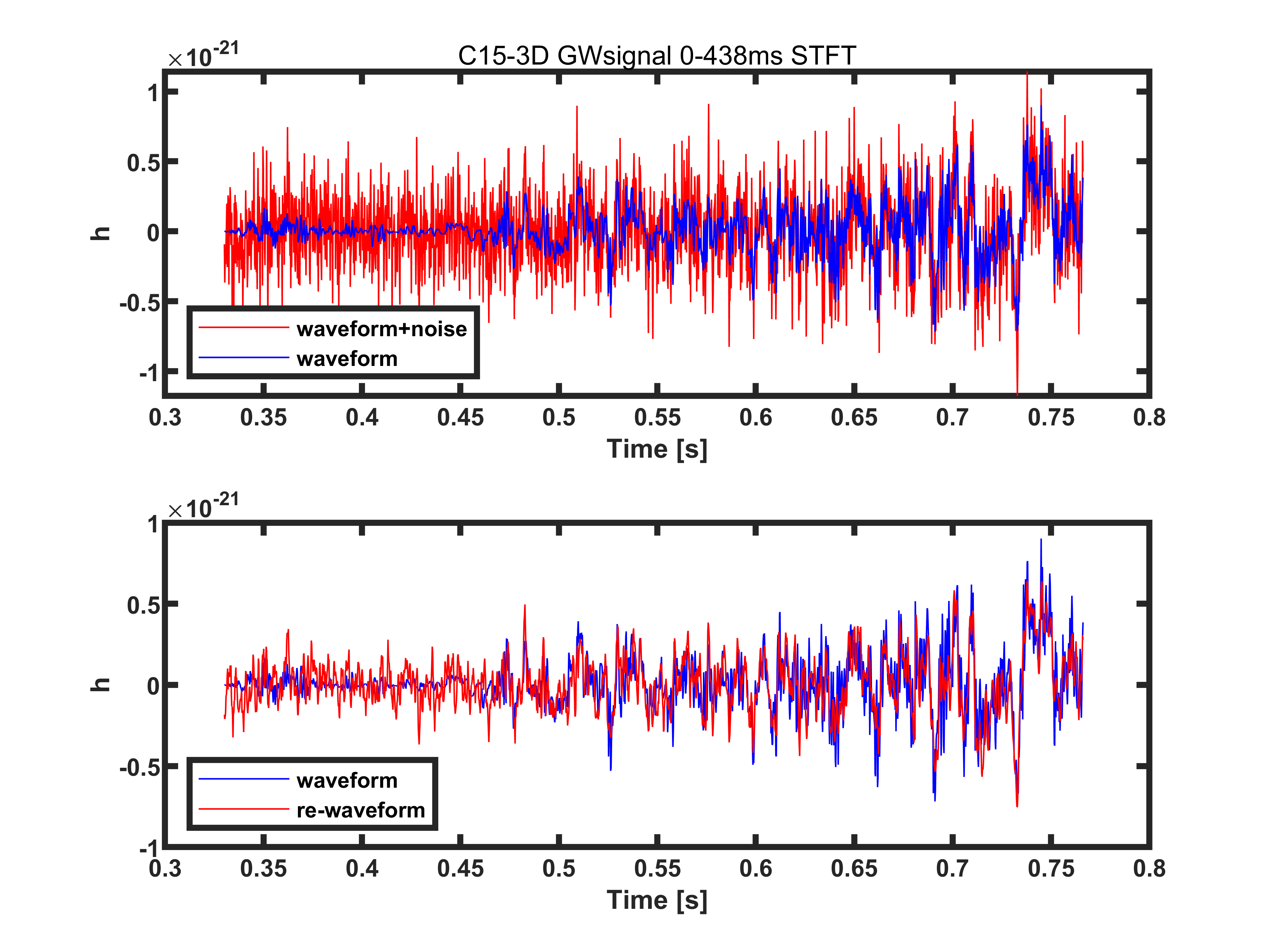}}
		\caption{These are the MSST, SET, and STFT results for data  C15-3D GWsignal 0-438ms in diffrent noise conditions. "Double the noise" means the noise amplitude is twice of the average amplitude of the signal.}\label{figx2}
	\end{figure}

\begin{figure}[htbp]
		\centering
		\subfigure[MSST result, triple the noise, SNR = 15.2527]{
			\includegraphics[width=0.45\textwidth]{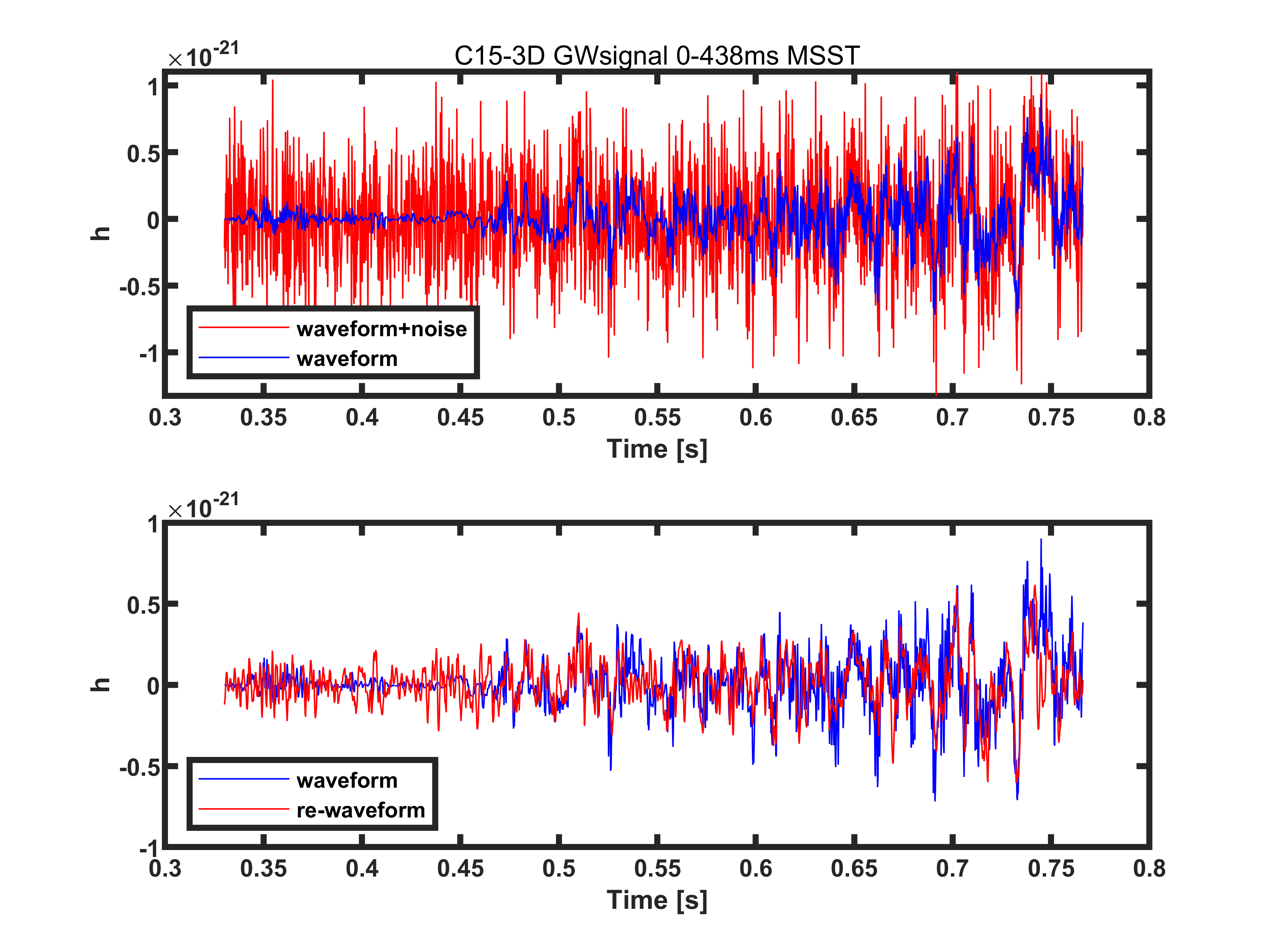}}
		\subfigure[SET result, triple the noise, SNR = 34.9722]{
			\includegraphics[width=0.45\textwidth]{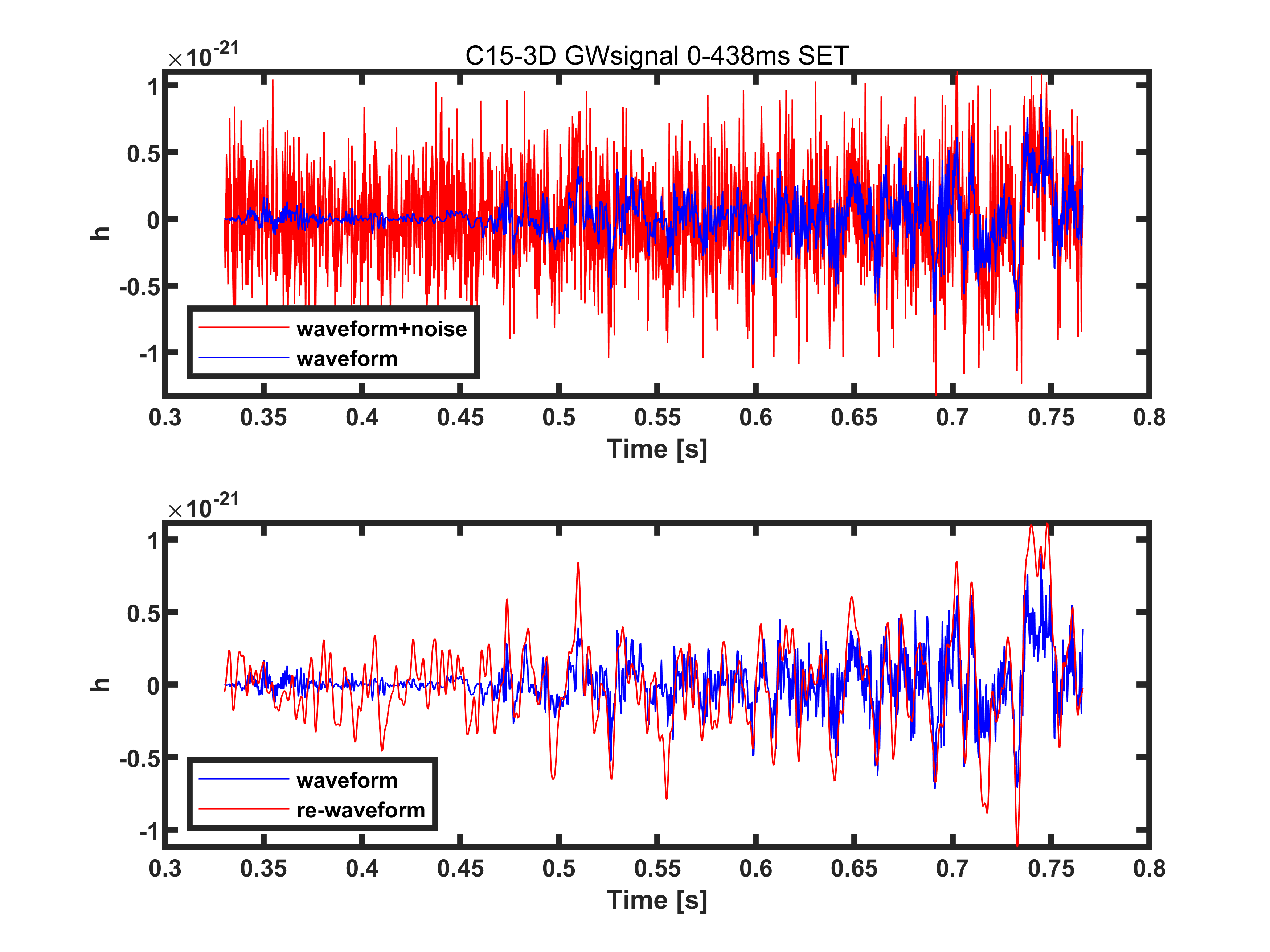}}
		\subfigure[STFT result, triple the noise, SNR = 26.7240]{
			\includegraphics[width=0.45\textwidth]{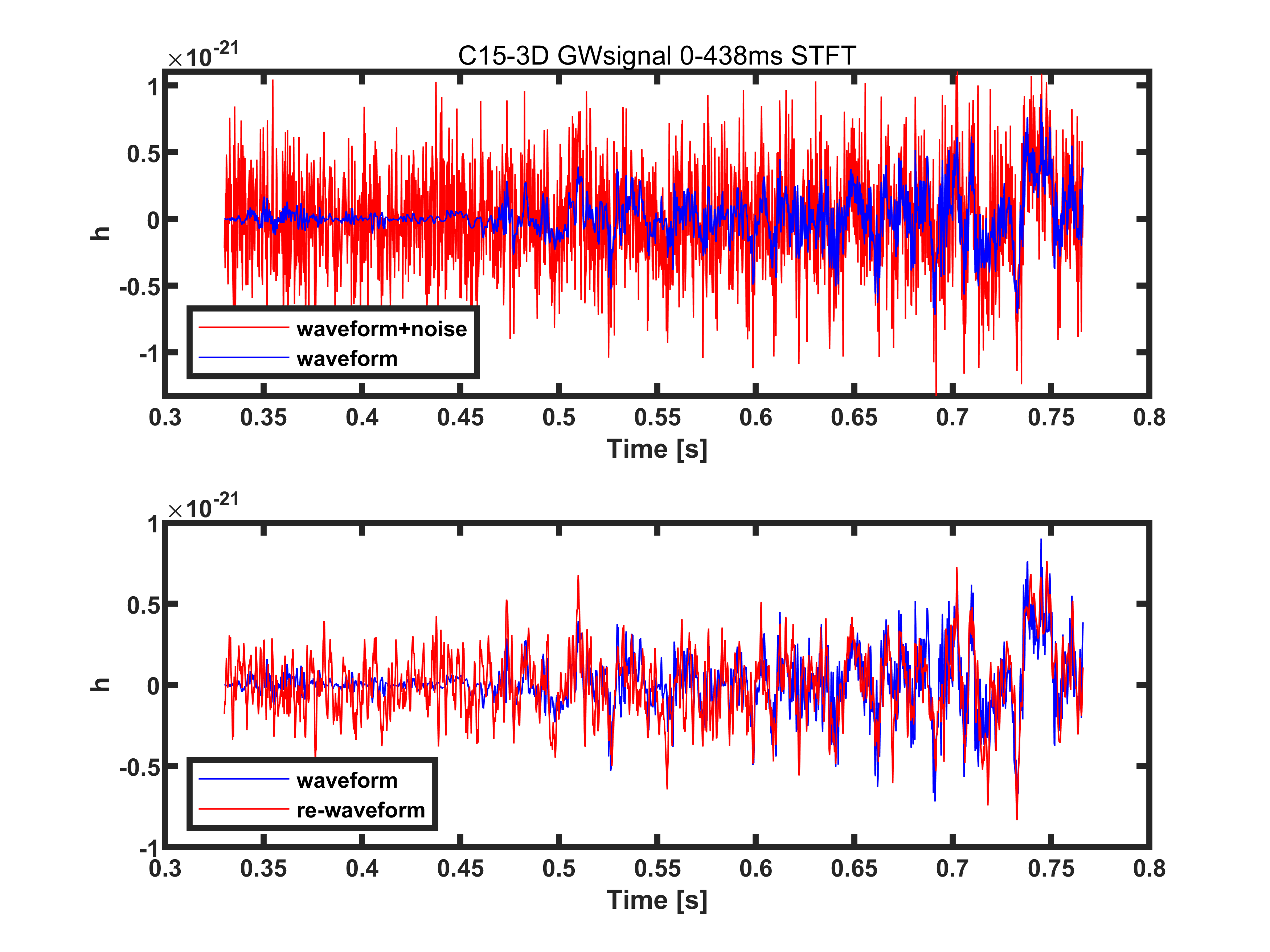}}
		\caption{Same as Fig.~\ref{figx2} but for  three times noise amplitude}\label{figx3}
		\end{figure}

\begin{figure}[htbp]
			\centering
		\subfigure[MSST result, five times the noise, SNR = 11.0244]{
			\includegraphics[width=0.45\textwidth]{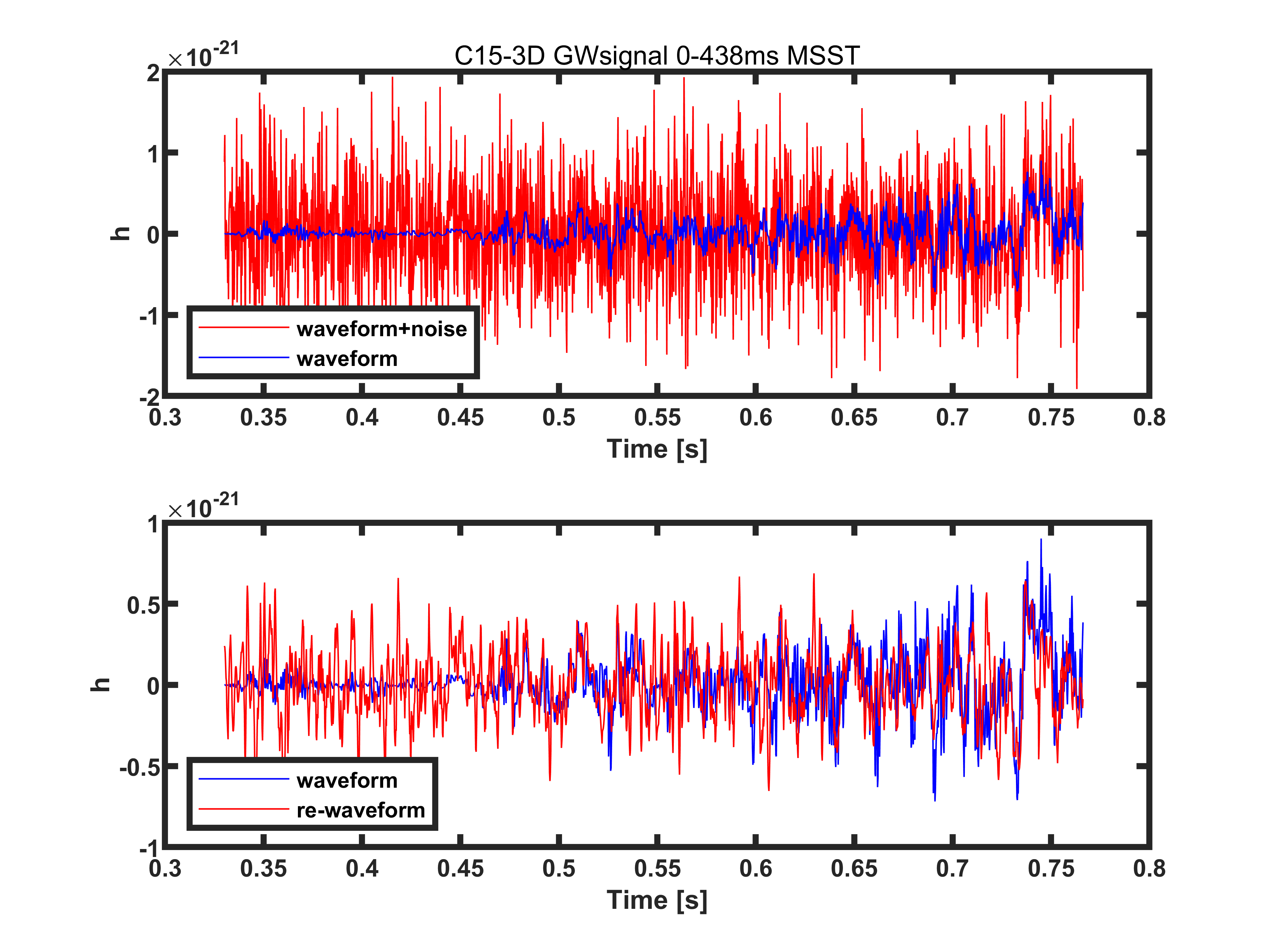}}
		\subfigure[SET result, five times the noise, SNR = 22.3214 ]{
			\includegraphics[width=0.45\textwidth]{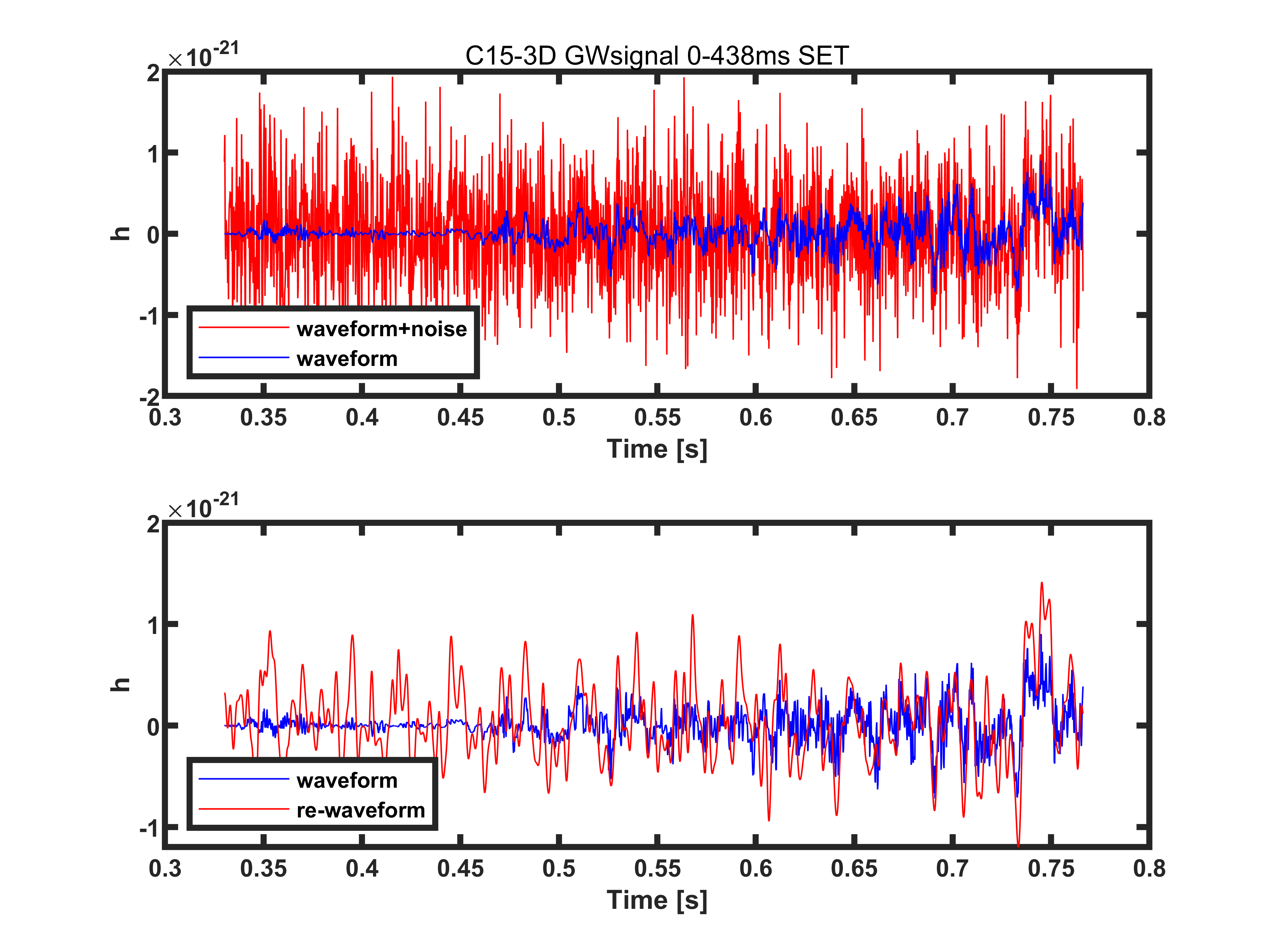}}
		\subfigure[STFT result, five times the noise, SNR = 20.8348]{
			\includegraphics[width=0.45\textwidth]{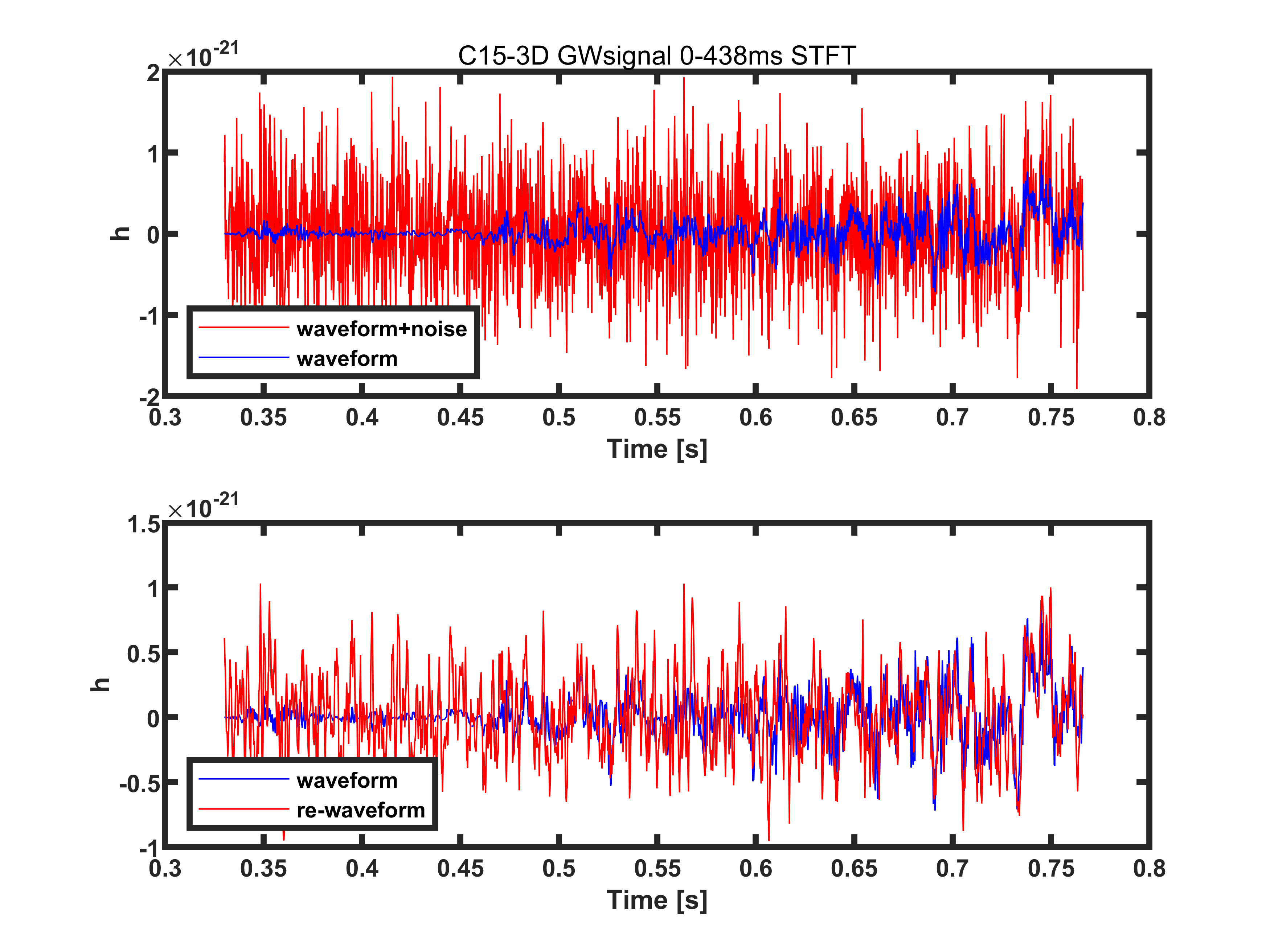}}
		\caption{Same as Fig.~\ref{figx2} but for  five times noise amplitude}\label{figx5}
	\end{figure}

	We further tested data with different  noise level than the last step, and the conclusion is that the pipeline is effective when the amplitude of the noise is approximately equal to or less than two times of the average amplitude of the signal. If noise is higher than that, the pipeline is unable to distinguish the noise and signal, facing the problem of reconstructing fake signals from noise. This is due to the high-frequency stochastic component of the SNGW signals, which is difficult to distinguish from the noise. The following figures show  results of waveform C15-3D GWsignal 0-438ms, we injected it into white noise whose average amplitude was the twice (Fig.~\ref{figx2}), three times (Fig.~\ref{figx3}), and five times Fig.~\ref{figx5} of the average amplitude of the signal.
		
		The effect is acceptable under double the amplitude of noise, but it is difficult to distinguish signals from noise. Under triple amplitude  noise, the SET method can roughly distinguish the mode, but the false signal caused by noise is gradually becoming obvious. The effect of five times noise is very poor. The false signal caused by noise cannot be ignored. In general, the SET method still performs the best.

\section{Conclusion}
We  applied three time-frequency methods, namely SET, MSST, and STFT, to the analysis of SNGW signals and tested the performances of them.  The tests were valued by the predicted SNR between the simulated data and the reconstructed signal. The noise we used was simulated whited noise.

Based on the results of our tests, we discovered that in terms of the time-frequency figure, SET and MSST were obviously better than STFT, which had great dispersion.
As for the performances of reconstruction, when valued by SNR, we discovered that the SET method performed the best and was efficient in finding the low-frequency mode, while the STFT method performed better than the MSST method. However, the STFT method would greatly increase the amount of clustering calculation in reconstruction because of dispersion.

Though the SET method seemed to be the best choice in our tests for gravitational wave signals, it is still too early to determine which time-frequency transform method is generally better or to which condition do they fit respectively since we only valued them with SNR, we will keep looking into this problem. We are planing to inject signals into real LIGO noise and calculate more statistics such as the ROC curve. We will also continue to improve our pipeline and explore more characteristics of these time-frequency transform methods in the next step of our research.

\textit{Acknowledgments.---}
The authors thank Jade Powell and Man Leong Chan for providing  the simulated waveforms.
We also  thanks Man Leong Chan and Yong Yan for valuable comments.

    \bibliographystyle{aasjournal}
    \bibliography{bibfile}
\end{document}